\documentclass[manuscript=article]{achemso}

\usepackage{chemformula} 
\usepackage[T1]{fontenc} 
\usepackage{graphicx}
\usepackage{dcolumn}
\usepackage{siunitx}
\usepackage{float}
\usepackage{bm}
\usepackage{amsmath}
\usepackage{mathtools}
\usepackage{placeins}
\usepackage{multirow}
\usepackage{booktabs}   
\usepackage{siunitx}    
\usepackage{caption}
\usepackage{amssymb}
\usepackage{afterpage} 
\usepackage{textgreek}
\usepackage[colorinlistoftodos]{todonotes}
\setkeys{acs}{biblabel=brackets,super=true}

\title{Perpendicular magnetic anisotropy in thin films enables extraordinary spin-wave phenomena: anti-Larmor precession, negative reflection and refraction, multi-reflection and multi-refraction
}
\author{Nikodem Leśniewski}
\email{nikles@amu.edu.pl}
    \affiliation{CNRS, Lab-STICC, UMR 6285, ENIB, 29238 Brest Cedex 3, France}
    \alsoaffiliation{Institute of Spintronics and Quantum Information, Faculty of Physics, Adam Mickiewicz University, Uniwersytetu Poznańskiego 2, 61-614 Poznań, Poland}
\author{Yuliya S. Dadoenkova}
    \affiliation{Université Jean Monnet Saint-Etienne, CNRS, Institut d’Optique Graduate School, Laboratoire Hubert Curien, UMR 5516, 42023 Saint-Etienne, France}
\author{Florian F. L. Bentivegna}
    \affiliation{CNRS, Lab-STICC, UMR 6285, ENIB, 29238 Brest Cedex 3, France}
\author{Paweł Gruszecki}%
     \email{gruszecki@amu.edu.pl}
    \affiliation{Institute of Spintronics and Quantum Information, Faculty of Physics, Adam Mickiewicz University, Uniwersytetu Poznańskiego 2, 61-614 Poznań, Poland}

\abbreviations{SW, PMA, DE, BV, FMR, SI, IFC}

\begin{document}


\date{\today}


\begin{abstract}
We present a theoretical and numerical investigation of the role of perpendicular magnetic anisotropy (PMA) in shaping spin-wave (SW) dynamics under low magnetic fields in thin and ultrathin magnetic films. PMA introduces an in-plane torque that counteracts exchange, dipolar, and Zeeman contributions, fundamentally modifying SW dispersion and inducing a local minimum that, under specific conditions, becomes the lowest frequency across all geometric configurations. This results in a sombrero-shaped dispersion in ultrathin films and a cowboy-hat-like shape in thicker films, where dipolar interactions dominate. Using isofrequency contour (IFC) analysis, we demonstrate that these PMA-induced dispersion shapes enable nontrivial wave phenomena unprecedented in uniform media: bireflection and negative reflection in ultrathin films, and trireflection in thicker films—where a single incident beam splits into three reflected components, two with negative angles. Most remarkably, we predict and demonstrate tri-refraction, where one incident beam generates three refracted beams with two exhibiting negative refraction angles. We further show anti-Larmor precession of magnetization near the dispersion minimum in thicker films, arising from the interplay between PMA-induced and dipolar torques. Systematic simulations across diverse material systems—metallic films, ferrimagnetic garnets, hybrid structures, and multilayers—confirm the universal nature of these phenomena in any PMA system supporting stripe domain transitions. These results open new opportunities to explore wave phenomena beyond magnonics.
\end{abstract}

\section{Keywords}

Spin waves, perpendicular magnetic anisotropy, thin film, specular reflection, precession

Refraction and specular reflection have been extensively studied across wave physics, revealing intriguing propagation phenomena.
The laws of reflection and refraction of visible light were first formulated in 984 by the Persian mathematician Ibn Sahl \cite{rashed1993}, rediscovered by Snell, and later formalized by Descartes in 1637 \cite{descartes1637} before being extended to any electromagnetic wave following Maxwell's contribution \cite{maxwell1865}. In its generalized form, Snell's law states that the tangential component of the wavevector to the interface is conserved during reflection and refraction--a principle known as phase matching. This fundamental rule governs waves in both isotropic and anisotropic media, and applies not only to electromagnetic waves \cite{hecht2016} but also, among others, to spin waves (SWs) \cite{gruszecki2014, stigloher2016, yu2016, mulkers2018, gruszecki2018, zhu2022}, which exhibit far more complex and tunable dispersion relations than light.

However, the development of metamaterials in recent decades has shown that the tangential component of the wavevector is not always conserved at an interface, leading to anomalous refraction and reflection phenomena \cite{Yu2014, Ni2012, Pfeiffer2013, Sun2012}. Among such effects for electromagnetic waves, negative refraction has attracted significant attention \cite{Poddubny2013, Macdo2014, Zhang2022, Grardin2019, Bang2019, jaksic2006, Zhang2009}. It is typically achieved \textit{via} strong anisotropy in the medium \cite{Macdo2014, kim2008, hioki2020, Zhang2022}, photonic metamaterials \cite{Zhang2009, Poddubny2013, Bang2019}, or specially engineered interfaces (\textit{e.g.}, metasurfaces introducing phase gradients) \cite{fleury2014, liu2017, mieszczak2020, zelent2019}.
In negative reflection of an electromagnetic wave from an interface, in contrast to negative refraction and anomalous reflection \cite{Wang2018}, the reflected wave appears on the same side of the normal to the interface as the incident wave; although the reflection angle is different from the angle of incidence, \textit{i.e.}, the reflected beam does not overlap the incident one. This effect has been less explored, with demonstrations limited to specific interface properties \cite{Meirbekova2023, Grardin2019, liu2018} or medium anisotropy \cite{Zhang2022, alvarez2022}.
Another notable non-specular effect is bireflection \cite{hioki2020, dadoenkova2022}, akin to birefringence, when an incident wave beam gives rise to two reflected beams. 
However, neither negative reflection nor bireflection of electromagnetic waves has been demonstrated in isotropic media at uniform interfaces without structural modifications such as diffraction gratings or metasurfaces.
In this article, we describe how negative and multiple reflections, along with other remarkable behaviors, can take place for SWs in thin magnetic films with perpendicular magnetic anisotropy.

In magnetic films, perpendicular magnetic anisotropy (PMA) favors the alignment of magnetic moments perpendicular to the film's surface \cite{stancil,hubert2008,KumarMishra2024}, affecting both the static magnetic configuration (\textit{e.g.} leading to periodic stripe domains at magnetic fields lower than the critical field $H^\mathrm{cr}$ \cite{hubert2008,fallarino2019}) and magnetization dynamics \cite{vukadinovic2000, Sucheta2019, grassi2022,  alvarez2021, Janardhanan2024}.
At still low magnetic fields but above $H^\mathrm{cr}$, in uniformly in-plane magnetized films, PMA creates a local minimum in the dispersion of Damon-Eshbach (DE) SWs \cite{banerjee2017} which, under specific conditions, can become a global minimum of the dispersion relation (with a frequency lower than the minimum for backward volume (BV) SWs) \cite{kisielewski2023}. 
A SW mode can be referred to as \textit{soft} when its frequency approaches zero as a function of some tuning parameter, such as magnetic field \cite{yoshizawa1985, Steiner1996}, wavevector \cite{kisielewski2023}, or temperature \cite{Sun2006}. This signals a critical point or phase instability, indicating a phase transition. The dynamics of such softened SW modes \cite{grassi2022} with frequencies approaching zero have been related to the periodicity and spatial distribution of the magnetization texture after the phase transition, just below $H^\mathrm{cr}$ from the uniform to the stripe-domain phase in films \cite{kisielewski2023} and stripes \cite{bailleul2003, Leaf2006}. Near this transition at $H^\mathrm{cr}$, a softened mode splits into a zero-frequency Goldstone mode and a non-zero-frequency Higgs mode on the low-symmetry side \cite{grassi2022}.
Although PMA suggests a variety of intriguing phenomena in SW dynamics near the phase transition, the behavior of SWs at low fields just above $H^\mathrm{cr}$ remains unexplored.

In this article, we investigate the influence of PMA on the dynamics of SWs in the linear regime in uniformly in-plane magnetized thin films at low magnetic fields near $H^\mathrm{cr}$. We demonstrate that PMA significantly modifies the overall torque acting on magnetization, leading to fundamental changes enabling the manifestation of novel phenomena. Our analysis begins with the simple case of an ultrathin film (with thicknesses up to several nanometers for which SWs from the fundamental band are uniform throughout the film's thickness) and progresses to the more complex scenario of thicker films. We show that the PMA-induced sombrero-like and cowboy-hat-like dispersion relations enable extraordinary wave phenomena including negative reflection, bireflection, trireflection, negative refraction, and tri-refraction—the splitting of a single incident beam into three refracted beams with two exhibiting negative angles. Furthermore, we demonstrate anti-Larmor precession of magnetization in thicker films near the dispersion minimum. Importantly, we show that these phenomena are universal across diverse magnetic materials—from metallic films (Permalloy) and garnets (YIG and  its doped variants) to hybrid structures and multilayers—and can be experimentally accessed.

\section{Results and discussion}

\subsection{Theoretical description of SWs in an ultrathin film with PMA.}
In the following, we consider thin magnetic films with surfaces parallel to the $xy$-plane and exhibiting PMA along the $z$-axis of a Cartesian system of coordinates.
For an ultrathin film uniformly magnetized in-plane along the $y$-axis by an external field $H_0$, the dynamic dipolar magnetic field can be approximated to the form $\mathbf{h}^\mathrm{d} = [h_x^\mathrm{d},0,h_z^\mathrm{d}] = [-\xi(k d) \mathrm{sin}^2 (\theta) m_x, ~0, ~-(1-\xi(kd))m_z]$, 
where $m_x$ and $m_z$ are the dynamic components of magnetization, $d$ is the film thickness, $\xi(kd)=1-(1-e^{-|kd|})/|kd|$, and $\theta$ is the angle, parallel to the $xy$ plane, of SW propagation with respect to the bias magnetic field \cite{kalinikos1986, gruszecki2015}. In such a film, the linearized Landau-Lifshitz equation (neglecting damping) for magnetization dynamics is given as:

\vspace*{-0.5cm} 
\begin{align}
\partial_t m_x &= \big( \overbrace{\frac{H_0}{M_s}}^{ =\scalebox{1.1}{$\frac{\tau_{x}^{0}}{c m_z }$} }  \overbrace{+l_\mathrm{ex}^2 k^2}^{=\scalebox{1.1}{$\frac{\tau_{x}^{\mathrm{ex}}}{c m_z }$}}   \overbrace{-\frac{h_z^\mathrm{d}}{m_z}}^{=\scalebox{1.1}{$\frac{\tau_{x}^{\mathrm{d}}}{c m_z }$}} 
   \overbrace{-Q}^{=\scalebox{1.1}{$\frac{\tau_{x}^{\mathrm{PMA}}}{c m_z }$}}\big) c m_z, \nonumber \\
\partial_t m_z &= \big( \underbrace{-\frac{H_0}{M_s}}_{ =\scalebox{1.1}{$\frac{\tau_{z}^{0}}{c m_x }$} } \underbrace{-l_\mathrm{ex}^2 k^2}_{ =\scalebox{1.1}{$\frac{\tau_{z}^{\mathrm{ex}}}{c m_x }$} } \underbrace{+  \frac{h_x^\mathrm{d}}{m_x} }_{ =\scalebox{1.1}{$\frac{\tau_{z}^{\mathrm{d}}}{c m_x }$}  } \big) c m_x
\label{eq:linLLE_v2}
\end{align}

where
$M_\mathrm{s}$ is the saturation magnetization,
 $l_\mathrm{ex}=\sqrt{A/(\frac{1}{2} \mu_0 M_\mathrm{s}^2) }$
is the exchange length and $A$ is the exchange constant, $\mu_0$ is the permeability of vacuum, $k$ is the absolute value of the wavevector, $Q=K_\mathrm{u}/(\frac{1}{2} \mu_0 M_\mathrm{s}^2 )$ is the reduced perpendicular magnetic anisotropy constant (in which $K_\mathrm{u}$ is the uniaxial anisotropy constant), and $c=|\gamma|\mu_0 M_\mathrm{s}$ where $\gamma$ is the gyromagnetic ratio. Note that the sign of $c$ defines the sense of precession of the ferromagnetic resonance (FMR) mode, which is the Larmor precession direction.

\begin{figure*}[!t]
    \vspace*{-1.1cm} 
    \centering
    \includegraphics[keepaspectratio,width=15cm]{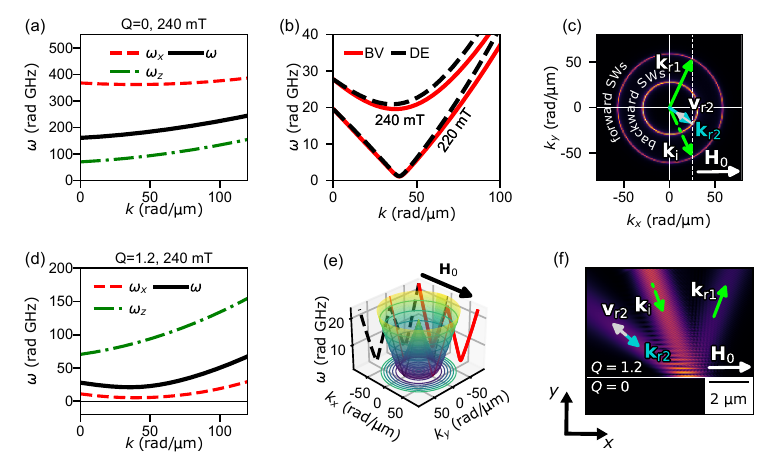}
    \vspace*{-0.5cm} 
    \caption{(a) and (d) SW dispersion for a $2$-nm-thick CoFeB film magnetized along the $y$-axis with $Q = 0$ (a) and $Q = 1.2$ (d): $\omega = \sqrt{\omega_x \omega_z}$ (black), $\omega_x$ (red, equation~(\ref{eq:wx})), and $\omega_z$ (green, equation~(\ref{eq:wz})).
    (b) Dispersion for external magnetic fields of $220$ mT (lower pair) and $240$ mT (upper pair) in both BV (red) and DE (dashed black) geometries.
    (e) Dispersion as a function of $k_x$ and $k_y$ for ultrathin film uniformly magnetized along the $x$-axis; red and black lines show the cross-sections for $k_y = 0$ (BV configuration) and $k_x = 0$ (DE configuration), with circles indicating examples of IFCs for different frequencies.
    (c) IFC at $1.5$ GHz and the magnetic field of value 380 mT applied along the $x$-axis computed using \textsc{mumax3}  showing incident and reflected wavevectors $\mathbf{k_\mathrm{i}}$, $\mathbf{k_\mathrm{r1}}$, and $\mathbf{k}_\mathrm{r2}$; the group velocity direction for $\mathbf{k_\mathrm{r2}}$ is indicated by the white arrow.
    (f) Micromagnetic simulation of the reflection of an SW beam in a 2 nm CoFeB film, incident at a $25^\circ$ angle on an interface separating regions with different $Q$ values.  
    Wavevectors (parallel to phase velocities) and group velocity directions from (e) are marked by arrows.
    }
    \label{fig:fig1}
    \vspace*{-0.5cm}
\end{figure*}

Equation ~\ref{eq:linLLE_v2} can be written as $[\partial_t m_x, \partial_t m_z] = [\tau^\mathrm{eff}_x, \tau^\mathrm{eff}_z]$, where $\tau^\mathrm{eff}_x$ and $\tau^\mathrm{eff}_z$ are the $x$- and $z$-components of the effective torque acting on magnetization. The total torque $\boldsymbol{\tau}^\mathrm{eff} = \boldsymbol{\tau}^0 + \boldsymbol{\tau}^\mathrm{ex} + \boldsymbol{\tau}^\mathrm{d} + \boldsymbol{\tau}^\mathrm{PMA}$ is the sum of terms representing torques due to the external (Zeeman), exchange, dipole, and PMA fields, respectively (see the definition of those torques in Supporting Information (SI)). PMA torque $\boldsymbol{\tau}^\mathrm{PMA} = c [-Q m_z,~0]$ has only an $x$-component and acts opposite to the other torques, squeezing the precession orbit along the $x$-axis.
If the PMA torque has a larger magnitude than the sum of all other torques, \textit{i.e.} if $Q > H_0/M_\mathrm{s} + l_\mathrm{ex}^2 k^2 + h_{z}^{\mathrm{d}}$ (for ultrathin films $-h_{z}^{\mathrm{d}}/m_z=(1-\xi(kd))>0$), the magnetic moment precession around the $y$-axis is expected to reverse. However, solving equation~(\ref{eq:linLLE_v2}) for this condition yields a time dependence of $m_{x,z}$ that is not harmonic ($\propto e^{-i\omega t}$) but instead grows unbounded over time (see SI). This occurs because the assumed uniform magnetic configuration along the $y$-axis is unstable in that case, leading to a phase transition to an out-of-plane stripe-domain configuration \cite{kisielewski2023}. 
It is worth noting that for the thin isotropic film, the direction of the material's magnetization in the $xy$ plane is irrelevant as long as the angle between the magnetization vector and the direction of SW propagation is preserved. We will revisit the reversal of the precession direction in the case of thicker films later in the manuscript.


\subsection{SW dispersion in ultrathin films with PMA.}

Let us first focus on the influence of PMA on SWs in ultrathin films. The dispersion profiles $\omega_x (k)$, $\omega_z (k)$ and $\omega (k) = \sqrt{\omega_x \omega_z}$ (see the Methods section for details on the definition of $\omega_x$ and $\omega_z$) for a 2 nm thick CoFeB film ($M_\mathrm{s} = 1344$ kA/m, $A = 13.6$ pJ/m) without PMA ($Q = 0$) and with PMA ($Q = 1.2$) are shown in Figure~\ref{fig:fig1}(a) and (d). One can see that $\omega_z$ increases monotonically with $k$, while $\omega_x$ exhibits a minimum corresponding to the $k$-vector for which exchange interactions start to dominate over dipolar interactions. The values of $Q$ and $H_0$ can shift $\omega_z$ up or down. For sufficiently high $Q$  and small $H_0$, the $\omega_x$ dependence can be shifted enough to introduce a minimum in $\omega$, with PMA playing a key role in this transition. The origin of this PMA-induced isotropy lies in the competition between anisotropic and isotropic contributions to the dispersion: the dipolar field component $h_x^{\mathrm{d}} = -\xi(kd)\sin^2(\theta)m_x$ introduces angular dependence through 
the $\sin^2(\theta)$ term in $\omega_z$, while PMA contributes an isotropic term $-Q$ to $\omega_x$ that is independent of propagation direction. When $Q$ is sufficiently large, this isotropic minimum in $\omega_x$ dominates over the anisotropic contribution in $\omega_z$, resulting in approximately isotropic dispersion $\omega = \sqrt{\omega_x\omega_z}$ at finite wavevector.

The effect of damping on the dispersion relation is also noteworthy. Although, 	in the analysis above, we neglected damping for simplicity, when damping $\alpha$ is included (see detailed derivation and analysis in SI), the dispersion relation can be written in the following approximate form that is amenable to analysis:

\begin{equation}
\mathrm{Re}(\omega) \approx \omega_0 
- \alpha^2\left[
    \omega_0 + \delta
\right], \label{eq:dispersionWithDamping_approx}
\end{equation}

where $\omega_0 = \sqrt{\omega_x \omega_z}$ and $\delta=(\omega_x-\omega_z)^2)/(8\omega_0) $. Equation~\eqref{eq:dispersionWithDamping_approx} shows that damping reduces the frequency proportionally to $\alpha^2$. However, in addition to the classical frequency reduction term $\alpha^2 \omega_0$~\cite{GurevichMelkov1996}, an additional contribution $\alpha^2 \delta$ emerges due to the difference between $\omega_x$ and $\omega_z$. While this term is typically negligible, it becomes significant during SW softening. As $\omega_x$ approaches zero, $\omega_0$ decreases correspondingly, causing $\delta$ to increase dramatically (since $\omega_z \neq 0$). Under these conditions, $\delta$ can exceed $\omega_0$ (see detailed analysis in the SI). 
Since this effect scales with $\alpha^2$, in materials with low damping that support SW propagation ($\alpha \lesssim 10^{-2}$), the damping-induced modifications are expected to be very small and observable only in the regime where the softened spin wave mode frequency approaches zero. Therefore, we will not focus on the impact of the damping in the subsequent parts of this paper.

Figure~\ref{fig:fig1}(b) shows the dispersion relations in a film with $Q = 1.2$ for SWs propagating along two orthogonal propagation directions (BV and DE configurations, for a propagation along and perpendicular to the bias magnetic field direction, respectively) in magnetic fields of 220 mT and 240 mT. A small decrease in the field significantly affects the dispersion: at 240 mT, the DE and BV SW dispersion minima occur around 20 rad GHz (with a slightly lower minimum frequency for the BV dispersion curve). At 220 mT, the minimum is reached for a near-zero frequency, with DE and BV dispersion curves overlapping for $k < 60$ rad/\textmu m.

The surface of the dispersion relation $\omega(k_x, k_y)$ shown in Figure~\ref{fig:fig1}(e) for a magnetic field of value $\mu_0 H_0=220$~mT applied along the $x$-axis resembles the sombrero potential that appears in various physical systems \cite{landau1980}. It features two circular isofrequency contours (IFCs) below the FMR frequency $f(k=0)$, an outer contour for forward SWs, where group and phase velocities are parallel, and an inner contour for backward SWs, where they are antiparallel. Figure~\ref{fig:fig1}(c) shows example contours for $f = 1.5$ GHz and bias magnetic field of value $\mu_0 H_0 =380$~mT applied along the $x$-axis, computed using \textsc{mumax3} (see the Methods section for details) \cite{Vansteenkiste2014}. 
Such a dispersion relation with a minimum extending isotropically for all angles of propagation opens new avenues in nonlinear physics, \textit{e.g.}, magnon Bose-Einstein condensation \cite{demokritov2006}.

The existence of those two IFCs implies that if the tangential component $k_x$  of the wavevector of the incident wave is smaller than the inner contour radius, two reflected waves (one forward and one backward) can form at a boundary between a magnetic film with PMA and another medium, as shown in Figure~\ref{fig:fig1}(c), where 
the wavevectors of one incident wave ($\mathbf{k_\mathrm{i}}$) and the corresponding reflected SWs ($\mathbf{k_\mathrm{r1}}$ for the forward SW, and $\mathbf{k_\mathrm{r2}}$ for the backward SW) are marked.
The forward wave follows the classically expected specular reflection rule, whereas the second reflected wave exhibits unique properties. 
First, its phase and group velocities are opposite, although it is not a classical backward volume magnetostatic mode (with $k_x=0$). Second, the energy of a reflected wave (associated with the direction of the group velocity) is expected to flow away from the boundary, with a positive $y$-component $v_{\mathrm{g},y}$ of the group velocity. However,  this is not the case for the backward SWs, whose wavevector lies in the same quadrant as that of the incident SW.

To verify this, we performed micromagnetic simulations (Figure~\ref{fig:fig1}(f); 
see the Methods section for details) of the reflection of a SW beam (with a frequency of 1.5~GHz) impinging under a 25$^\circ$ angle of incidence on a sharp interface separating two uniform magnetic regions. The first region, where the SWs are excited, exhibits PMA ($Q=1.2$), whereas the second region does not ($Q=0$). The magnetic field is applied along the $y$-axis. The simulation results indeed reveal two reflected beams, \textit{i.e.}, a bireflection. 
The first beam, with wavevector $\mathbf{k_\mathrm{r1}}$, is reflected under the same angle (in absolute value) as the angle of incidence and obeys the rules of specular reflection. 
 The second beam with wavevector $\mathbf{k_\mathrm{r2}}$ does not obey those rules.
 It exhibits a negative angle of reflection, \textit{i.e.,} the reflected beam is located on the same side of the normal as the incident wave. Moreover, the direction of its the group velocity is antiparallel to its wavevector.
This behavior corresponds to what is traditionally called negative reflection. It is here obtained for SWs, whereas it must be emphasized that it has not been reported for electromagnetic waves at a similar boundary between isotropic media. Indeed, while negative reflection has been reported for light \cite{Meirbekova2023}, but also for polaritons \cite{Zhang2022}, and acoustic waves \cite{Grardin2019}, these instances relied on a strong anisotropy in the dispersion relation of the medium or on the nonuniformity of the interface (\textit{e.g.}, diffraction gratings). To the best of our knowledge, negative reflection in a uniform, isotropic medium at a sharp interface has never been demonstrated for any type of wave so far.
Moreover, since the SW dispersion is isotropic here, this effect should be observed regardless of the angle between the field and the incident wavevector. This is unusual because for a layer without PMA, even a thin one, the dispersion exhibits anisotropy due to the angular dependence of the dipolar field contribution, causing frequency differences between DE and BV configurations.

\subsection{Thicker films with PMA} 

The distinction between ultrathin and ``thicker'' film regimes is determined by the uniformity of the magnetization profile across the film thickness: when $m_{x,z}(z)$ remain approximately constant (maintained by exchange interactions for $d \lesssim l_\mathrm{ex}$), the dipolar field (derived from the magnetostatic potential, see Methods) can be thickness-averaged, yielding the easy-to-follow analytical expression used above \cite{kalinikos1986}. For thicker films where $d \gtrsim l_\mathrm{ex}$, the SW profile may develop a $z$-dependence driven by dipolar interactions—particularly pronounced for Damon-Eshbach modes \cite{Damon_Eshbach}—and the validity of the thickness-averaged approximation diminishes gradually, necessitating comparison with numerically computed profiles \cite{rychly2016, Szulc2022}.

\begin{figure}
    \centering
    \includegraphics[keepaspectratio,width=16cm]{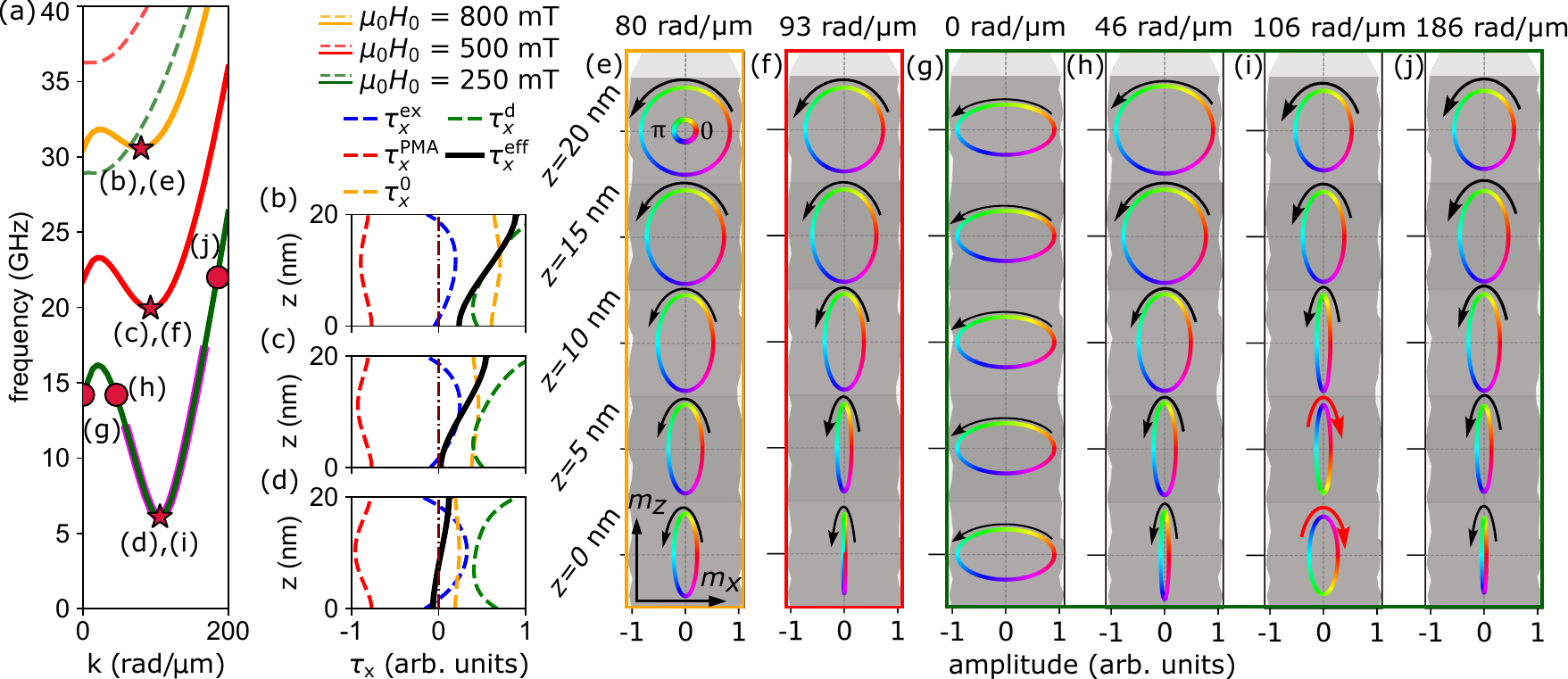}
        \vspace*{-.3cm} 
    \caption{(a) Dispersion of SWs propagating along the $x$-axis, originating from the first (bold solid lines) and second (narrow dashed lines) bands of a 20 nm CoFeB film with $Q = 0.6$, under a bias magnetic field $H$ applied along the $y$-axis with values of 250 mT (green), 500 mT (red), and 800 mT (orange). Letters (b)-(j) correspond to the $k$ and field values marked by dots and stars in (a). The thick magenta line for 250 mT shows the range where anti-Larmor precession occurs.
    (b)-(d) $x$-component of the torque ($\tau_x$) across the film thickness for wavevectors marked by stars in (a), corresponding to wavevectors and fields of $80$ rad/\textmu m at 800 mT, $93$ rad/\textmu m at 500 mT, and $106$ rad/\textmu m at 250 mT. The dashed blue, green, red, and orange lines show the exchange, dipolar, PMA, and Zeeman components of $\tau_x$, respectively, with the solid black line indicating the effective torque.
    (e)-(j) Precession orbits of the magnetic moments at $z = 0, 5, 10, 15,$ and $20$ nm. Colors represent phase, while black and red arrows indicate Larmor and anti-Larmor precession, respectively. Panel (e) corresponds to 800 mT, panel (f) to 500 mT, and panels (g-j) to 250 mT. $k$ values are shown at the top of each subplot. See SI for the animated version of (e)-(j).
    }     \label{fig2:elipsy}
\vspace*{-0.5cm}
\end{figure}

Having established this framework, we now test the hypothesis that increasing film thickness can lead to anti-Larmor precession in films with PMA. For the 20 nm thick CoFeB film with PMA ($Q = 0.6$), the $z$-dependent SW profile results in a concentration of the SW amplitude at one of the film's surfaces in the DE configuration \cite{Damon_Eshbach}. We therefore employ the finite-element method (FEM) in \textsc{COMSOL Multiphysics} \cite{rychly2016, Szulc2022, Szulc2024} to solve equation (\ref{eq:linLLE_v2}) coupled with Gauss's equation for the dipolar field, computing the dispersion relation and mode profiles (see Methods for details). The results for three values of the external magnetic field (800, 500, and 250 mT) applied along the $y$ axis are shown in Figure~\ref{fig2:elipsy}(a).

The dispersions for the DE configuration are nonmonotonic and exhibit both a local minimum and a local maximum as $k$ increases, similarly to the experimental observations reported \cite{banerjee2017,grassi2022,dhiman2024}. The initial positive slope towards the local maximum results from stronger dipole interactions in thicker films compared to the ultrathin case (\textit{cf.} Figure~\ref{fig:fig1}(b)). When $k$ increases further, the dispersion curve then exhibits a local minimum that becomes more pronounced and shifts towards larger wavevectors when the applied magnetic field decreases \cite{kisielewski2023}, which resembles the behavior observed in ultrathin films (\textit{cf.} Figure~\ref{fig:fig1}(b)).

Figures~\ref{fig2:elipsy}(e-j) depict the precession orbits at five positions (at $z=0,5,10,15$ and $20$~nm) across the film thickness for several different combinations of wavevector and magnetic field values that correspond to the (e)-(j) labels in Figure~\ref{fig2:elipsy}(a).
For all cases except the FMR mode (Figure~\ref{fig2:elipsy}(g), for $k=0$), the precession ellipticity varies with the $z$-coordinate.
In Figs.~\ref{fig2:elipsy}(e),(f),(h) and (j), the precession orbit becomes more polarized along the $z$-axis for small values of $z$ (\textit{i.e.}, close to the bottom film surface), since the amplitude of $m_x$ becomes reduced while approaching that surface.
In these cases, a normal anticlockwise Larmor precession is observed. However, a peculiar and key result of our study is shown in Figure~\ref{fig2:elipsy}(i), that shows the precession orbits for an applied field ($\mu_0 H_0=250$ mT) and an incident wavevenumber ($k_x=106$~rad/\textmu m) corresponding to the deepest minimum of the dispersion curves in Figure~\ref{fig2:elipsy}(a). Across the film thickness, there is a point where the precession direction of the magnetic moments is reversed (somewhere between $z=5$~nm and $z=10$~nm). This effect occurs only for wavevectors near the minimum of the dispersion curve, as marked with the magenta highlight of the branch corresponding to $\mu_0 H_0=250$~mT, in Figure~\ref{fig2:elipsy}(a).
The occurrence of such an anti-Larmor precession near a magnetic film surface was predicted theoretically \cite{salanskii1975, grunberg1982}. In Ref.~\citenum{dieterle2019}, these modes were termed heterosymmetric SWs (due to the different symmetry of $m_x$ and $m_z$) and attributed to the hybridization between the fundamental and first perpendicular standing SW modes (second dispersion band) \cite{dieterle2019, trevillian2024, heins2025}. 
In our system, the frequency difference between the first and second SW bands (dashed lines in Figure~\ref{fig2:elipsy}(a)) is on the order of dozens of GHz, \textit{i.e.}, 14.7 GHz at $k=0$, and for wavevectors exhibiting anti-Larmor precession above 20 GHz (27 GHz at $100$ rad/\textmu m), as can be seen in Figure~\ref{fig2:elipsy}(a). 
Such a large frequency gap makes the interaction between the first and second SW bands unlikely, suggesting a different origin for these modes.
To explain the origin of the heterosymmetric profile of the SWs with anti-Larmor precession and confirm that they result from the presence of PMA in the magnetic film, we analyzed how all the torque terms acting on magnetization from equation (\ref{eq:linLLE_v2}) change through the film thickness. The analysis for the $x$-component of the torques and SWs from the dispersions minimum (see Figure~\ref{fig2:elipsy}(a)) for three different values of $\mu_0 H_0$, namely 800, 500, and 250 mT is shown in Figure~\ref{fig2:elipsy}(b), (c), and (d), respectively.
A more detailed analysis of the $x$- and $z$-components of the torques can be found in the SI. 
The PMA-related torque $\tau_{x}^\mathrm{PMA}$ has a sign generally opposite to that of the other torques, as well as a large amplitude, and can thus counteract their influence if it can lead to a sign reversal of the total torque $\tau_{x}^\mathrm{eff}$. However, such a reversal does not take place for the larger values of the applied field, nor close to the upper surface of the film. Indeed, due to the varying strength of dipolar interactions across the film, the dipolar torque component $\tau_{x}^{\mathrm{d}}$, that does not depend much on the magnetic field, is stronger near that surface, leading to a larger $\tau_{x}^{\mathrm{eff}}$ there. 
As the bias field decreases, on the other hand, the Zeeman torque component $\tau_{x}^{\mathrm{0}}$ weakens, causing $\tau_{x}^{\mathrm{eff}}$ to decrease. As for the exchange torque component $\tau_{x}^\mathrm{ex}$, it is small near the film boundaries for all values of the bias field. The consequence of these dependences is that for smaller values of the field (here, at 250 mT) and close to the bottom surface of the film, the overall torque component  $\tau_{x}^{\mathrm{eff}}$ can cross zero, which can reverse the precession direction, as seen in Figure~\ref{fig2:elipsy}(i). Note that such a reversal also requires that the $z$- component $\tau_{z}^{\mathrm{eff}}$ of the total torque does not cross zero, which is indeed the case, as shown in the SI.
This clearly indicates that we have identified a new class of heterosymmetric SWs, where anti-Larmor precession near the bottom surface of the magnetic film results from a PMA-induced torque exerted on the magnetic moments, but also from a nonuniform dipolar field across the film, typical for DE SWs in thicker films.

\begin{figure*}[t!]
    \vspace*{0.cm} 
    \centering
    \includegraphics[keepaspectratio,width=15cm]{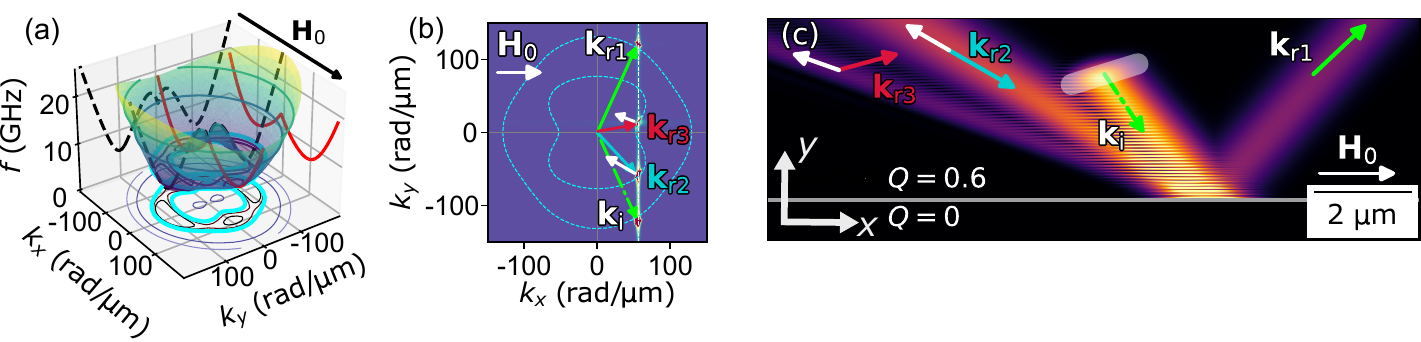}
    \vspace*{-0.4cm} 
    \caption{
    (a) Dispersion $f(k_x, k_y)$ for a 20 nm CoFeB film magnetized along the $x$-axis ($Q = 0.6$, $H_0 = 250$ mT), calculated  using the finite element method. The red and black curves show cross-sections for $k_x = 0$ (BV) and $k_y = 0$ (DE) configurations. Contours at the bottom represent IFCs at different frequencies, with the $9$~GHz contour highlighted in cyan.
    (b) IFC at 9 GHz (cyan dashed lines), extracted from (a). The colormap in the background (representing the 2D Fourier transform of (c)) shows small bright spots (see SI for a larger version of the colormap). 
    (c) Reflection of an SW beam in a 20 nm CoFeB film, incident at $24.5^\circ$ from the $Q = 0.6$  region to the $Q = 0$ region at a sharp interface. The semi-transparent white area marks the excitation region.  
    Wavevector directions in (b) and (c) are shown by green ($\mathbf{k_\mathrm{i}}$, $\mathbf{k_\mathrm{r1}}$), cyan ($\mathbf{k_\mathrm{r2}}$), and red ($\mathbf{k_\mathrm{r3}}$) arrows, with white arrows indicating group velocity directions for nonspecular reflections associated to $\mathbf{k_\mathrm{r2}}$ and $\mathbf{k_\mathrm{r3}}$.
    }     \label{fig:optics_thick}
\vspace*{-0.5cm}
\end{figure*}

\begin{figure}[hbt!]
    \centering
    \includegraphics[width=15cm]{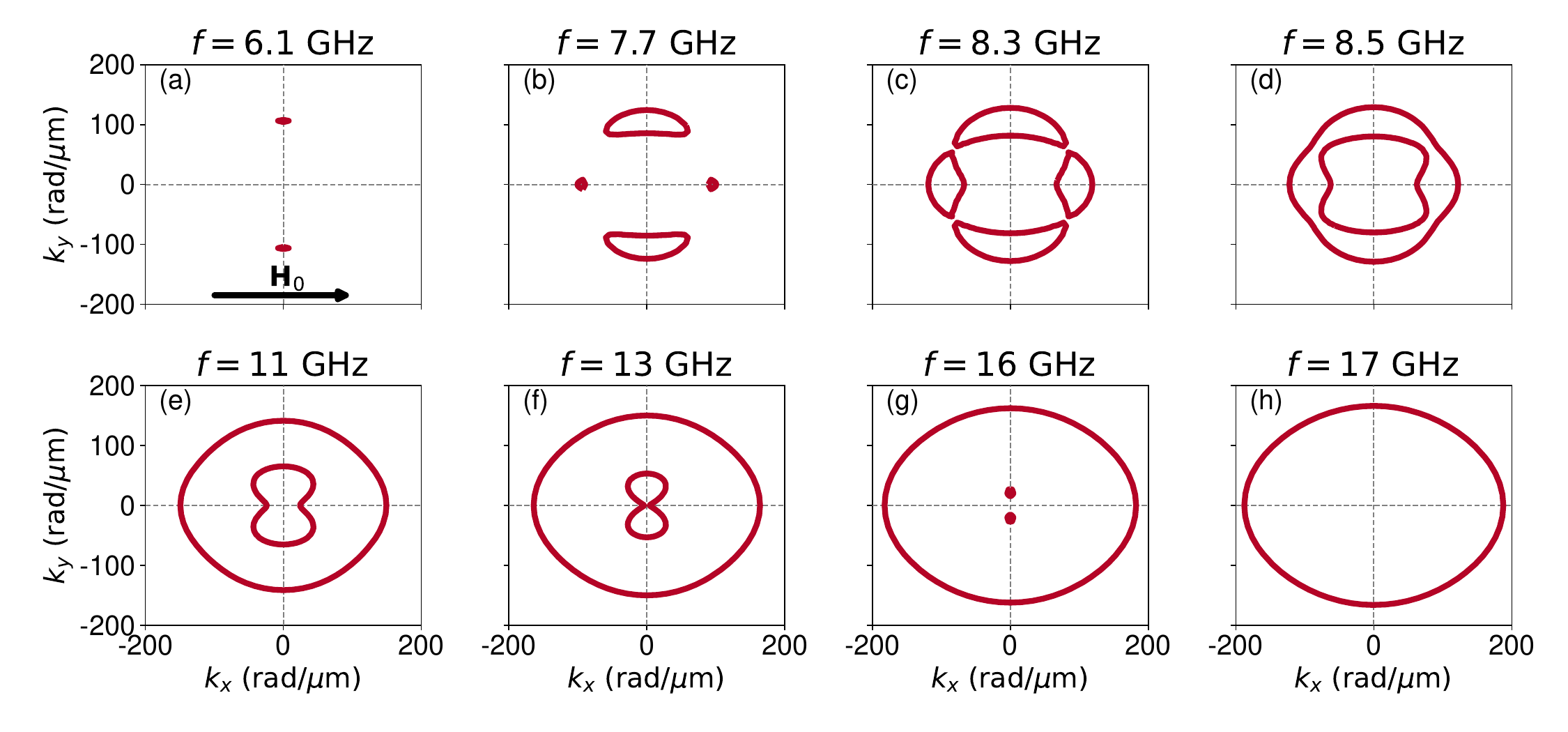}
    \caption{IFCs extracted from the dispersion relation presented in Figure3 for (a) $6.1$~GHz, (b) $7.7$~GHz, (c) $8.3$~GHz, (d) $8.5$~GHz, (e) $11$~GHz, (f) $13$~GHz, (g) $16$~GHz, and (h) $17$~GHz calculated for SWs in a 20 nm-thin CoFeB film with PMA ($Q=0.6$, $\mu_0 H_0=250$~mT). The arrow in (a) shows the direction of the external magnetic field $H_0$ in all the subsequent plots. An animated version of this plot can be found in the SI.
    }
    \label{fig:ifc}
\end{figure}

Next, we examine how increased dipole interactions, resulting from a greater film thickness, affect the sombrero-like shape of the dispersion relation observed for an ultrathin film. 
Again, we reorient the magnetic field direction towards the $x$-axis to clarify the visualization and analysis of oblique SW reflection that will be provided later in this section.
As shown in Figure~\ref{fig:optics_thick}(a), the dispersion for a 20 nm thick CoFeB film with $Q=0.6$ and the magnetic field applied of value 250 mT is anisotropic, transitioning from a sombrero to a cowboy-hat shape.
This change introduces a greater variety of IFCs (shown on the bottom surface of the plot), including frequencies with two closed contours (\textit{e.g.}, at 9 GHz, cyan lines) or even four closed contours (\textit{e.g.}, at 7.9 GHz, black lines). 
Figure~\ref{fig:ifc} illustrates this variety of IFCs for a set of frequencies ($6.1$~GHz, $7.7$~GHz, $8.3$~GHz, $8.5$~GHz, $11$~GHz, $13$~GHz, $16$~GHz, and $17$~GHz).
It is worth highlighting the unusual shapes of IFCs presented in Figure~\ref{fig:ifc}(a-c) and (g). Because of the frequencies chosen near the dispersion relation extrema, in Fig~\ref{fig:ifc}(a), (b), and (g), one can see isolated dots associated with peaks and dips of the function corresponding to SWs with zero group velocity. 
In Figure~\ref{fig:ifc}(c), the inner and outer contours are split into four separate closed curves.
Quasi-flat segments are also visible in the IFCs shown in Figure~\ref{fig:ifc}(b-f), which are essential for observing self-caustics\cite{veerakumar2006, kim2016, heussner2018}.
All of the presented shapes are prominent candidates for further investigation within the framework of SW reflection and refraction analysis, hinting at the rich potential of physical phenomena that can be observed in a homogeneous system with such a complex dispersion relation.
In the next paragraph, we will demonstrate an effect that illustrates this potential.

\subsection{Reflection of spin waves in thicker films with PMA} 
Focusing on the 9 GHz IFC represented by cyan curves at the bottom surface of Figure~\ref{fig:optics_thick}(a) and in Figure~\ref{fig:optics_thick}(b), we observe that the outer and inner contours correspond to forward (group velocity directed away from the IFC center) and backward (group velocity directed toward the IFC center) SWs. As these contours are no longer circular, the direction of the group velocity can differ significantly from that of the phase velocity, unlike in ultrathin layers, where these velocities are either parallel or antiparallel.

As an example indicating the rich potential of such complex IFCs, let us demonstrate the \textit{tri-reflection} of SWs (a scenario in which three reflected waves with different wavevectors can be observed). 
As presented in Figure~\ref{fig:optics_thick}(b), this can, for instance, be achieved with an incident wavevector  $\mathbf{k_\mathrm{i}}$ corresponding to the angle of incidence $24.5^\circ$ for which one can find three possible wavevectors for the reflected waves (with the same value of $k_x$ and a positive $y$-component of the group velocity $v_{\mathrm{g},y}$).

To verify this prediction, we performed micromagnetic simulations with a SW beam incident at $24.5^\circ$ on the sharp interface separating a film with PMA ($Q=0.6$) from a film without PMA, $Q=0$--see Figure~\ref{fig:optics_thick}(c) and details of simulations in the Methods section. Three reflected SW beams are indeed observed: a specular one with the same angle as the incident beam (wavevector $\mathbf{k_\mathrm{r1}}$), and two with negative reflection angles ($-60^\circ$ for wavevector $\mathbf{k_\mathrm{r2}}$ and $-70.7^\circ$ for wavevector $\mathbf{k_\mathrm{r3}}$). The two-dimensional spatial Fourier transform of this simulation, shown in the background of Figure~\ref{fig:optics_thick}(b), reveals four bright spots on the IFC towards which the arrows are pointing, corresponding to the same $k_x$ value and the predicted wavevectors. The second reflected wave, with wavevector $\mathbf{k_\mathrm{r2}}$, has similar properties to the negatively reflected wave demonstrated for the ultra-thin film (Figure~\ref{fig:fig1}(f)), with its group velocity direction opposite to its wavevector. The third reflected wave, with wavevector $\mathbf{k_\mathrm{r3}}$, lies on the inner IFC in the same quadrant as the ordinary specular reflected wave (wavevector $\mathbf{k_\mathrm{r1}}$) on the outer IFC. This wave can propagate outwards from the interface due to the anisotropic IFC, allowing the group velocity to be directed outward from the boundary (\textit{i.e.}, with $v_{\mathrm{g},y} > 0$), with the group velocity nearly perpendicular to the wavevector. Additionally, the third beam is narrower and exhibits almost no spreading, characteristic of caustic-like beams whose wavevectors lie on flat segments of the IFC (as in the case of $\mathbf{k_\mathrm{r3}}$). To the best of our knowledge, this is the first numerical observation of tri-reflection of any kind of wave at a uniform plane boundary between two media, without nanostructuring, complex interface engineering, or modulation of the beam itself. This finding may provide new insights into wave propagation and applications involving multiple-wave interactions at interfaces.
It should be noted that, whatever the number of reflections, the negatively reflected beams presented here are always accompanied by a specularly reflected beam, which is not always the case for the anomalous reflection of electromagnetic waves at inhomogeneous interfaces, where negative reflection can be observed while no specular reflection takes place \cite{Macdo2014}.

\begin{figure}
    \vspace*{0.cm} 
    \centering
    \includegraphics[width=8.6cm]{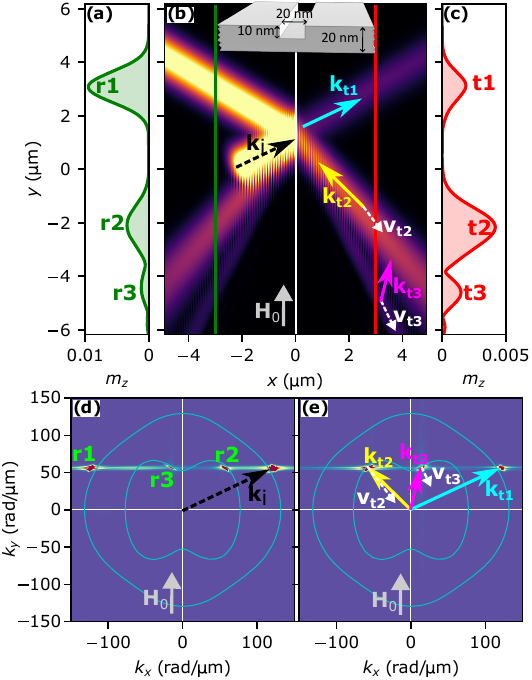}
    \caption{
    Reflection and refraction of a SW beam at frequency 9 GHz in a 20 nm thick CoFeB film with $Q=0.6$, incident at $24.5^\circ$ at a 10-nm-deep and 20-nm-wide groove (see schematic representation at top of (b)). The magnetic field $H_0$ of 250 mT is applied along the interface ($y$-axis).
\textbf{(b)} Steady-state spatial distribution of $|m_z|$ showing one incident beam with wave vector $\mathbf{k}_\mathrm{i}$, three reflected beams ($r1$, $r2$, $r3$) with wave vectors $\mathbf{k}_\mathrm{r1}$, $\mathbf{k}_\mathrm{r2}$, $\mathbf{k}_\mathrm{r3}$ (scenario discussed in Fig.~\ref{fig:optics_thick}), and three transmitted (refracted) beams ($t1$, $t2$, $t3$) with wave vectors $\mathbf{k}_\mathrm{t1}$, $\mathbf{k}_\mathrm{t2}$, and $\mathbf{k}_\mathrm{t3}$. The groove interface is marked by a white line at $x=0$.
\textbf{(a, c)} $|m_z|$ profiles as a function of $y$ at $x = -3$~μm (a, incident beam region with beams labeled $r1$, $r2$, $r3$) and at $x = 3$~μm (c, transmitted beam region with beams labeled $t1$, $t2$, $t3$).
\textbf{(d, e)} Two-dimensional Fourier transforms of the magnetization distribution from panel b for the regions where $x < 0$ (d, incident/reflected region) and $x > 0$ (e, transmitted region). Cyan contours represent isofrequency curves at 9 GHz. Arrows indicate the wave vectors of the incident beam ($\mathbf{k}_\mathrm{i}$) and transmitted beams ($\mathbf{k}_\mathrm{t1}$, $\mathbf{k}_\mathrm{t2}$, $\mathbf{k}_\mathrm{t3}$). White arrows show the group velocity directions $\mathbf{v}_\mathrm{t2}$ and $\mathbf{v}_\mathrm{t3}$ (normal to the isofrequency contours at the corresponding wave vector values), indicating the directions of energy transfer that in these cases do not coincide with the corresponding wavevector direction. 
    }     \label{fig:triRefraction}
\vspace*{-0.5cm}
\end{figure}

\subsection{Negative refraction and tri-refraction of spin waves in thicker films with PMA} 
Having demonstrated the rich physics of SW reflection in films with PMA, let us now investigate how SW refraction manifests itself in this system under the same field regime. 
The simplest experimental implementation for studying both reflection and refraction of SWs is a groove geometry, which offers remarkable experimental simplicity. It requires only a homogeneous magnetic material with a single groove etched using standard fabrication techniques such as focused ion beam milling. This straightforward approach eliminates the need for complex multilayer structures or compositional gradients, making it highly accessible for experimental implementation.
Figure~\ref{fig:triRefraction} illustrates the complex reflection and refraction phenomena of a spin wave beam incident at 24.5° on a 10-nm-deep groove in a 20 nm thick CoFeB film at 9 GHz. The spatial distribution of the out-of-plane magnetization component in Figure~\ref{fig:triRefraction}(b) reveals a rich multimode behavior: a single incident beam splits into three reflected beams ($r1$, $r2$, $r3$—with two exhibiting negative angles of reflection as discussed previously) and three transmitted beams ($t1$, $t2$, $t3$), each characterized by distinct wave vectors.

The three reflected and refracted beams with their respective amplitudes are clearly visible in the cross-sectional views taken 3~μm away from the interface on either side, as shown in Figure~\ref{fig:triRefraction}(a) and (c). For the reflected waves, we observe the same scenario as previously discussed in Fig.~\ref{fig:optics_thick}. The transmitted spin waves also exhibit three refracted modes, analogous to the reflection case. Notably, the largest amplitude corresponds to beam $t2$, which refracts at a negative angle.

To understand these observations, Figure~\ref{fig:triRefraction}(d) and (e) present the data from panel (b) in $k$-space, separately for the $x<0$ (incident/reflected) and $x>0$ (transmitted) regions. All bright spots are located on the isofrequency contours at 9 GHz. The reflected region shows exactly the same pattern as in Fig.~\ref{fig:optics_thick}(b). For the transmitted spin waves, we observe three bright spots with identical $k_y$ values, corresponding to the wavevectors of spin waves excited at the interface: $\mathbf{k}_\mathrm{t1}$, which matches the incident wavevector $\mathbf{k}_\mathrm{i}$; $\mathbf{k}_\mathrm{t2}$, located in the opposite quadrant of $k$-space on the inner isofrequency contour to $\mathbf{k}_\mathrm{i}$; and $\mathbf{k}_\mathrm{t3}$, positioned on the inner contour in the same quadrant of $k$-space as the incident spin wave.

To understand the propagation directions of the refracted beams, particularly for $t2$ and $t3$, one must analyze the group velocity directions. The group velocity for a given wave vector $\mathbf{k}_\mathrm{t}$ is normal to the isofrequency contour surface at that wave vector. As shown by the white arrows in Figure~\ref{fig:triRefraction}(e), the group velocities $\mathbf{v}_\mathrm{t2}$ and $\mathbf{v}_\mathrm{t3}$ (marked by white arrows in Figure~\ref{fig:triRefraction}(b,e)) do not align with their corresponding wave vectors, revealing the strongly anisotropic nature of the SW dispersion relation. The refracted beam $t2$ lies on the inner IFC in the opposite quadrant, with its group velocity direction opposite to its wavevector, similar to the negatively reflected beam $r2$ demonstrated in Fig.~\ref{fig:optics_thick}. The third refracted beam $t3$ lies on the inner IFC in the same quadrant as the specularly refracted beam $t1$ on the outer IFC. This wave can propagate forward across the interface due to the anisotropic IFC, allowing the group velocity to be directed away from the boundary (\textit{i.e.}, with $v_{\mathrm{g},x} > 0$), with the group velocity nearly perpendicular to the wavevector.


\subsection{Material universality and experimental accessibility}

To the best of our knowledge, this is the first observation of tri-refraction--the splitting of a single incident beam into three refracted beams with two exhibiting negative angles of refraction--for any type of wave. The experimental simplicity of this geometry, requiring only a linear groove fabricated using standard techniques, makes this system highly accessible for experimental verification. Importantly, the dominant refracted beam $t2$ exhibits a negative refraction angle, and its wavevector is located on the inner isofrequency contour corresponding to longer wavelengths. This configuration can be made experimentally feasible to verify using standard detection techniques such as micro-focused Brillouin light scattering (\textmu BLS) or time-resolved magneto-optical Kerr effect (TR-MOKE) by appropriately selecting the operating frequency to ensure that the inner isofrequency contour lies within the measurable range of these techniques ($k < 20$ rad/μm). Examples of materials that support SWs with such small 	wavevectors can be Yttrium Iron Garnet (YIG) Y$_3$Fe$_5$O$_{12}$ \cite{prestwood2025}, a ferrimagnet with minimal damping, or YIG with the substitution of materials like Bismuth Bi:YIG \cite{das2024} or Cerium Ce:YIG \cite{ghising2017} on the Yttrium sites. Furthermore, the phenomena described in this work can be observed for a wide variety of materials with PMA that, granted the proper thickness and bias field, host a transition between uniform magnetization and a stripe domain pattern. The Table \ref{tab:material_params} presents a collection of materials and composites with their corresponding parameters for which such a transition has been documented. 

Up to this point, the analysis in the manuscript has focused on 2 nm and 20 nm thick layers with CoFeB material parameters and varying PMA strengths ($Q=0$, $1.2$ for the 2 nm layer and $Q=0.6$ for the 20 nm layer). However, the influence of PMA on the dispersion relation and the PMA-induced spin wave softening process described in this work is universal across a broad range of materials. To demonstrate this universality, we performed a series of FEM simulations for selected experimentally studied materials. Following the relationship between spin waves and stripe domain patterns established in Ref.~\cite{kisielewski2023}, we selected materials based on their ability to support a stripe domain configuration in the remanent state.

\begin{figure}[ht!]
    \centering
    \includegraphics[width=16cm]{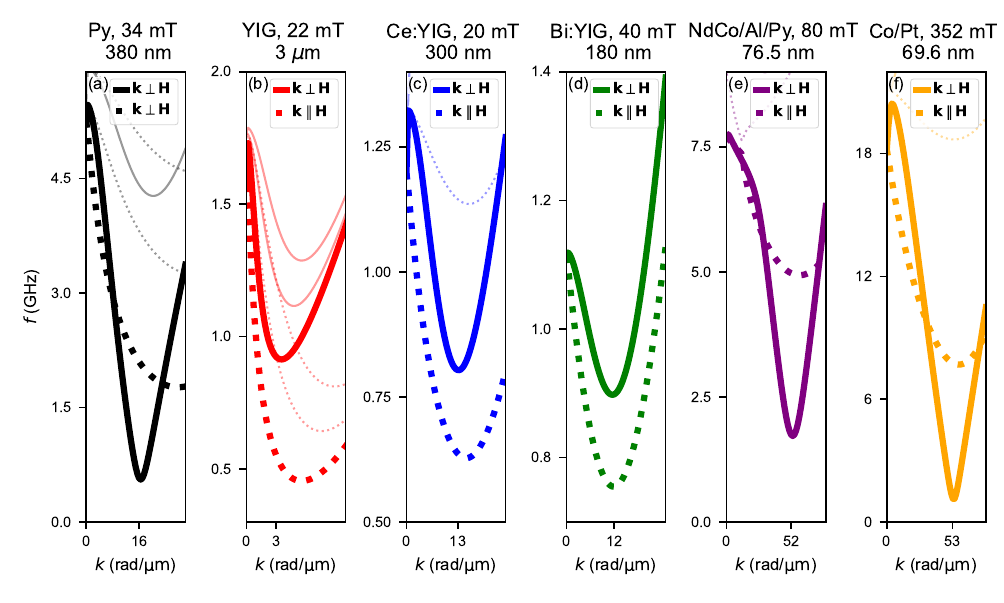}
    \caption{Dispersion curves calculated for in-plane external field directed parallel (dotted lines) and perpendicular (solid lines) to the SW propagation calculated for a collection of materials in which a stripe-domain pattern of magnetization has been observed. Table \ref{tab:material_params} holds the parameters for each material. Bold lines and dots signify the fundamental frequency mode; thin and pale lines and dots signify higher eigenfrequencies. On the $x$-axis, an approximate wavevector $k$, for which the minimum frequency in DE geometry is present, has been marked. }     \label{fig:materials}
\end{figure}

The material parameters of the selected systems are presented in Table~\ref{tab:material_params}. For each material, we systematically varied the bias field to identify field values at which a clear minimum emerges in the dispersion relation for spin waves propagating perpendicular to the in-plane applied bias magnetic field. The resulting dispersion relations are shown in Fig.~\ref{fig:materials}, with the corresponding external magnetic field value indicated in each panel title.

\begin{table}
\centering
\small
\caption{Material parameters of selected experimentally reported systems with PMA that according to our analysis exhibit mode softening and resulting dispersion relations with sombrero and cowboy-hat shapes.}
\label{tab:material_params}
\begin{tabular}{llccccc}
\toprule
Paper & Material & $M_s$ (kA/m) & $A$ (pJ/m) & $Q$ & $\alpha$ & $d$ (nm) \\
\midrule
Voltan et al. (2016) \cite{voltan2016} & Py & 859 & 13.00 & 0.05 & $5\cdot 10^{-3}$ & 380 \\
\cmidrule(lr){1-7}
Prestwood et al. (2025) \cite{prestwood2025} & YIG & 139 & 6.50 & 0.11 & $5\cdot 10^{-5}$ & 3000 \\
\cmidrule(lr){1-7}
Ghising et al. (2017) \cite{ghising2017} & Ce:YIG & 81 & 1.20 & 0.28 & $10^{-4}$ & 300 \\
\cmidrule(lr){1-7}
Das et al. (2024) \cite{das2024} & Bi:YIG & 40 & 0.79 & 1.00 & $10^{-4}$ & 180 \\
\cmidrule(lr){1-7}
\multirow{3}{*}{\shortstack[l]{Szulc et al. (2022) \cite{Szulc2022}\\NdCo/Al/Py trilayer}} & NdCo$_{7.5}$ & 1100 & 10 & 0.17 & 0.1 & 64 \\
& Al spacer & --- & --- & --- & --- & 2.5 \\
& Py & 860 & 10 & 0\textsuperscript{a} & $10^{-2}$ & 10 \\
\cmidrule(lr){1-7}
Dhiman et al. (2024) \cite{dhiman2024} & Co/Pt & 1237.5 & 26.25\textsuperscript{b} & 0.60 & $10^{-2}$ & 69.6 \\
\bottomrule
\multicolumn{7}{l}{
\begin{tabular}[t]{@{}l@{}}
    \textsuperscript{a} Calculated using $K_\mathrm{IMA} = 1.2$~kJ/m$^3$ (in-plane anisotropy).\\
    \textsuperscript{b} Different exchange stiffness in the perpendicular direction: $A_z=19.7$ pJ/m.
  \end{tabular}
}
\end{tabular}
\end{table}

In Fig.~\ref{fig:materials}, solid lines represent the DE geometry with the bias field perpendicular to the spin wave propagation direction, while dotted lines represent the BV geometry with the bias field parallel to the propagation direction. Bold thick lines correspond to the fundamental frequency band, and pale thin lines represent higher-order modes.

Fig.~\ref{fig:materials}(a) shows the dispersion relation for a 380 nm thick Permalloy (Py, Fe$_{20}$Ni$_{80}$) film~\cite{voltan2016} with very low PMA in an external field of 34 mT. However, due to large thickness and increased role of dipolar interactions, a clear minimum is present at $k \approx 16$~rad/$\mu$m. Figs.~\ref{fig:materials}(b)-(d) present results for garnet-based materials: (b) 3~$\mu$m thick YIG~\cite{prestwood2025} in 22 mT field, (c) 300 nm thick Ce:YIG~\cite{ghising2017} in 20 mT  field, and (d) 180 nm thick Bi:YIG~\cite{das2024} in 40 mT  field. The dispersion minimum appears at small wavevectors for all YIG-based materials, with pure YIG exhibiting the smallest value at $k \approx 3$~rad/$\mu$m.

Fig.~\ref{fig:materials}(e) shows a hybrid trilayer system NdCo(64)/Al(2.5)/Py(10)~\cite{Szulc2022}, where the numbers in parentheses indicate layer thicknesses in nanometers. The system consists of a 64 nm thick NdCo$_{7.5}$ layer (the subscript denotes 7.5 at.\% Co composition) with high PMA and damping, separated from a 10 nm thick Py layer (low damping and in-plane anisotropy) by a 2.5 nm thick Al insulating spacer. The material parameters for both magnetic layers are listed in Table~\ref{tab:material_params}. Both layers share a gyromagnetic ratio of $\gamma = 185$~$\mathrm{rad\cdot GHz/T}$. The NdCo layer has $K_\mathrm{PMA} = 130$~kJ/m$^3$ and in-plane magnetic anisotropy $K_\mathrm{IMA} = 10$~kJ/m$^3$, while the Py layer has $K_\mathrm{IMA} = 1.2$~kJ/m$^3$. The dispersion minimum occurs at $k \approx 52$~rad/$\mu$m. This system demonstrates how PMA characteristics can be imprinted from a high-damping PMA layer onto a low-damping film without intrinsic PMA, producing softening behavior analogous to that described earlier in the manuscript.

Finally, Fig.~\ref{fig:materials}(f) presents results for a Co/Pt multilayer~\cite{dhiman2024} with structure \\ Ti(4)/Pt(30)/[Co(2.2)/Pt(0.7)]$_{24}$/Pt(2.3), where the bracket notation indicates 24 repetitions of the Co(2.2)/Pt(0.7) bilayer. The gyromagnetic ratio is $\gamma = 190$~$\mathrm{rad\cdot GHz/T}$. Such Co/Pt multilayer structures are widely used to achieve thick films with strong PMA, as the perpendicular magnetic anisotropy originates from interfacial effects at each Co/Pt interface and increases with the number of bilayers. This approach enables the fabrication of relatively thick magnetic layers (approximately 70 nm in total) while maintaining substantial effective PMA. The dispersion minimum is present at $k \approx 53$~rad/$\mu$m.

These results demonstrate that the PMA-induced spin wave softening phenomenon is not limited to CoFeB but is a general feature observed across diverse magnetic materials—from metallic films (Py) and garnets (YIG and its doped variants) to hybrid structures and multilayers. Despite differences in material parameters spanning orders of magnitude, all systems exhibit the characteristic dispersion minimum when appropriate PMA and bias field conditions are satisfied, confirming the universal nature of this effect.

Finally, we want to discuss the conditions for an experimental observation of bi- and tri-reflection, as well as negative reflection of SWs. A key limitation of standard optical measurement methods, such as micro-focused Brillouin light scattering or time-resolved magneto-optical Kerr effect, is their ability to detect SWs with wavelengths below 500-600 nm \cite{sebastian2015, perzlmaier2008}. However, Mie-resonance-enhanced micro-focused Brillouin light scattering can measure SWs with wavelengths as short as 125 nm \cite{wojewoda2023}, and X-ray microscopy is expected to measure SWs below 100 nm \cite{grafe2019}. Recently, soft-X-ray momentum microscopy has been demonstrated to directly measure magnon dispersions and interactions at wavelengths below 100 nm~\cite{wittrock2025}, enabling simultaneous access to both short-wavelength modes on the outer isofrequency contour and their longer-wavelength counterparts on the inner contour. This technique would be particularly well-suited to a comprehensive characterization of tri-reflection and negative reflection phenomena at high wavevectors, as it can directly resolve all three reflected beams with their distinct wavelengths in momentum space without requiring wavelength-selective detection.

Interestingly, since two IFCs exist for certain frequencies, the wavelength of non-specularly reflected SWs (on the inner IFC) is much longer than that of the incident and specularly reflected SWs. Therefore, with the right antenna to excite short-wavelength obliquely incident SWs (that cannot be measured using micro-focused Brillouin light scattering or time-resolved magneto-optical Kerr effect), it should be possible to detect SWs reflected at a negative angle even using standard techniques.
\textit{E.g.}, at frequency 13 GHz with the IFC shown in Figure~\ref{fig:ifc}(f), an excited incident wave with a wavelength of approximately 41 nm will produce a negatively reflected wave with a wavelength of 565 nm, well within the detection capabilities of such techniques.

The experimental accessibility of these phenomena can be significantly enhanced by appropriate material selection. Among the materials presented in Table~\ref{tab:material_params}, garnet-based systems with PMA—particularly YIG and its doped variants (Ce:YIG, Bi:YIG)—offer substantial advantages for experimental verification. Due to their larger exchange length ($l_\mathrm{ex} \approx 16$ nm for YIG compared to $\approx 3.7$ nm for CoFeB), these materials exhibit the characteristic dispersion minimum at much smaller wavevectors. For instance, pure YIG shows a minimum at $k \approx 3$~rad/$\mu$m (corresponding to a wavelength of $\approx 2~\mu$m), well within the detection range of standard micro-focused Brillouin light scattering or time-resolved magneto-optical Kerr effect techniques. This is in stark contrast to CoFeB films, where the minimum occurs at $k > 60$~rad/$\mu$m ($\lambda < 100$~nm), requiring advanced detection methods. Furthermore, YIG-based materials possess exceptionally low Gilbert damping ($\alpha \sim 10^{-5}$ for pure YIG, $\alpha \sim 10^{-4}$ for doped variants), enabling SW propagation over millimeter-scale distances and facilitating the observation of reflection and refraction phenomena over macroscopic length scales. The combination of accessible wavelengths, low damping, and experimentally demonstrated stripe domain formation~\cite{prestwood2025, ghising2017, das2024} makes garnet-based systems with PMA ideal candidates for the first experimental demonstrations of the phenomena described in this work.

It is important to note that the existence of two isofrequency contours at certain frequencies introduces an additional experimental consideration: conventional broadband antenna excitation can simultaneously excite mode pairs with wave vectors $k$ and $-k'$ that have opposite phase velocities but identical signs of group velocity. This leads to interference patterns in the spatial distribution of magnetization dynamics, resulting in position-dependent precession characteristics and oscillating SW amplitudes even at distances far from the excitation region (see detailed analysis in SI). For quantitative studies of bi- and tri-reflection phenomena, this necessitates either narrowband excitation schemes to selectively address individual modes or explicit modeling of multi-mode interference effects in the interpretation of experimental data.

\section{Conclusions}

We have presented a theoretical and numerical investigation of spin wave dynamics in thin magnetic films with perpendicular magnetic anisotropy at low fields, revealing unprecedented wave phenomena with implications for wave physics beyond magnonics.

Our central finding is that PMA fundamentally transforms the torque landscape acting on magnetization, enabling two remarkable effects. First, in thicker films, we demonstrate anti-Larmor precession arising from the interplay between PMA-induced torque and the nonuniform dipolar field characteristic of Damon-Eshbach modes—a mechanism distinct from previously reported mode hybridization. Second, PMA dramatically reshapes the spin wave dispersion relation: ultrathin films exhibit an isotropic sombrero-like dispersion, while thicker films display an anisotropic cowboy-hat-like form. In both cases, frequency ranges emerge where forward and backward spin waves coexist across all propagation directions, creating inner and outer isofrequency contours.

These dispersion characteristics enable wave phenomena not previously observed for any wave type in uniform media at sharp interfaces: negative reflection and bireflection in ultrathin films, and trireflection in thicker films where a single beam splits into three reflected components. Most remarkably, we demonstrate tri-refraction at a simple groove interface—one incident beam generating three refracted beams with two at negative angles. These effects arise naturally from the isofrequency contour topology, without requiring metamaterials or interface engineering.

These phenomena are universal across magnetic thin films with PMA. Systematic simulations across diverse materials—metallic films, ferrimagnetic garnets, hybrid trilayers, and multilayers—confirm that all systems exhibiting stripe domain transitions display the characteristic dispersion minima associated with these highly peculiar features. Material selection critically impacts experimental accessibility: garnet-based systems (YIG and doped variants) with larger exchange lengths exhibit minima at wavevectors  $\sim$20 times smaller than metallic films, yielding wavelengths ($\approx 2~\mu$m) accessible to conventional detection techniques. Combined with ultralow damping ($\alpha \sim 10^{-5}$), these systems provide ideal experimental platforms. Advanced soft X-ray momentum microscopy further enables sub-100 nm characterization. The dual-contour topology introduces a caveat: broadband excitation generates interfering mode pairs, necessitating narrowband schemes or interference modeling for quantitative studies.

The dynamics of softened SW modes near the dispersion minimum in PMA films shows great promise for nonlinear physics applications, as the response of softened SWs to microwave excitation is enhanced. These findings highlight the potential to engineer dispersion relations through the interplay of PMA and magnetic-field-induced torques, opening new opportunities to explore wave phenomena beyond magnonics and into broader 
wave physics.


\section{Methods}

\subsection{Simulation details}


\subsubsection{FEM-based eigenspectrum calculation in COMSOL Multiphysics (eigenfrequency study)}

To compute the dispersion relation, mode profiles, and precession orbits of a $20$ nm thick CoFeB film shown in Figure~\ref{fig2:elipsy} and Figure~\ref{fig:optics_thick}(a-b), we have solved the linearized Landau-Lifshitz equation for SW propagation along the $x$-axis using the finite-element method in \textsc{COMSOL Multiphysics} \cite{Szulc2021, Szulc2022, Szulc2024}.
The linearized Landau-Lifshitz equation takes the form
\begin{align}
    i \omega m_x &= -|\gamma|\mu_0\left( \left[H_0 - M_\mathrm{s}Q - \frac{2 A}{\mu_0 M_\mathrm{s}} \Delta \right] m_z + M_\mathrm{s} \partial_{z} \varphi \right),\\
    i \omega m_z &= |\gamma|\mu_0 \left( \left[H_0 - \frac{2 A}{\mu_0 M_\mathrm{s}} \Delta \right] m_x + M_\mathrm{s} \partial_{x} \varphi \cdot \cos(\theta) \right),
\end{align}
where $\theta$ is the angle of the magnetic field with respect to the $y$ axis ($\theta=\pi/4$ for DE configuration), $\varphi$ is the magnetostatic potential that can be computed from the magnetostatic Maxwell equations written in the form of a Poisson-like equation:

\begin{equation}
    \Delta \varphi - \partial_x m_x \cdot \cos(\theta) - \partial_z m_z=0.
    \label{eq:poisson}
\end{equation}

The modeled thin film is surrounded above and below by a non-magnetic medium that can be taken as a vacuum without loss of generality. Dirichlet boundary conditions are applied to the top and bottom surfaces of the film to enforce a zero magnetostatic potential. At the lateral sides (left and right edges of the system), Floquet boundary conditions are used to simulate the behavior of an infinite film.
We used the standard triangular mesh with a maximum element size of $1$~nm in the magnetic material. The study is performed with the use of an eigenfrequency solver with a Parametric Sweep of the wavevector values in the range $0-200$~rad/\textmu m, and a Parametric Sweep of the angle $\theta$.

\subsubsection{Micromagnetic simulations}

The results from Figure~\ref{fig:fig1}(e-f) and Figure~\ref{fig:optics_thick}(b-c), presented as color maps, are obtained from micromagnetic simulations using \textsc{mumax3} \cite{Vansteenkiste2014}, which solves the time-dependent Landau-Lifshitz-Gilbert equation via the finite-difference method. The simulations and postprocessing methods follow \cite{Gruszecki2022, Sobucki2023}. The system, a CoFeB film magnetized along the $x$-axis, has thicknesses of 2 nm (380 mT field) and 20 nm (250 mT field). The dimensions of the system are $L_x \times L_y \times L_z$, discretized with cells $l_x \times l_y = 5 \times 5$ nm$^2$, and $l_z = 2$ nm or $\frac{20}{7}$ nm for the 2-nm and 20-nm films, respectively. The equilibrium static configuration is obtained first, followed by dynamic simulations where SWs are excited by a time-dependent magnetic field. Absorbing boundary layers of width 1200 nm are used to prevent edge reflections.

\subsubsection{Point-source excitation and calculation of isofrequency contours}

For ultra-thin films, the SWs are excited by a microwave field from a point source at the center of the $xy$ plane, across the whole thickness of the sample:
\begin{equation}
    b_{\mathrm{PS}}(t, x, y) = A_\mathrm{exc} G(x, y) \sin(2 \pi f_0 t) (1-\exp(-0.002 \pi f_0 t),
\end{equation}
where $A_\mathrm{exc} = 0.005B_0$, $f_0 = 1.5$ GHz, and $G(x, y) = \exp\left(-\frac{x^2 + y^2}{2\sigma^2}\right)$ with $\sigma = 20$ nm. The SWs are excited for 1.5 ns to reach the steady state. After recording 80 snapshots, we compute the pointwise FFT of the $m_z$ component over time. Then we select only results for the excitation frequency $f_0$ and apply a 2D FFT over $x$ and $y$ and plot the absolute value of the FFT to derive the IFCs, as shown in Figure~\ref{fig:fig1}(e) for the 2-nm thick CoFeB film.

\subsubsection{Excitation of an obliquely incident spin-wave beam}

For an obliquely propagating SW beam, the microwave field is given by:
\begin{equation}
\begin{split}
    b_\mathrm{beam}(t;x, y) = A_\mathrm{exc}(1-e^{-0.002\pi f_0 t})G_x(x)G_y(y)\times \\ \times [\sin(k_0 x)\sin(2\pi f_0 t)+\cos(k_0 x)\cos(2\pi f_0 t)],
\end{split}
\end{equation}
where $A_\mathrm{exc} = 1$ $\mu$T, and $k_0$ is the wavevector modulus. Gaussian functions $G_\xi(\xi) = \exp\left[-\xi^2/(2 \sigma_\xi^2)\right]$, $\xi=x,y$, define the profile of the beam, with $\sigma_x = 5\lambda$, $\sigma_y = 5\lambda$ for 2 nm thick, and $\sigma_x = 5\lambda$, $\sigma_y = 10\lambda$ for 20 nm thick films. The direction of emission of the field is rotated to produce an SW beam incident at an angle $\theta$. We excite SWs until steady-state, then we record 80 snapshots, perform a pointwise FFT over time, and select only results for the excitation frequency $f_0$. To represent the SW amplitude in the $k$-space, we perform a 2D FFT over $x$ and $y$ and plot its absolute value as the colormap.

\subsubsection{Bireflection of spin waves in a 2-nm-thick CoFeB film}

For the study of bireflection (Figure~\ref{fig:fig1}(f)), we simulate a system with dimensions $7620 \times 3072 \times 2$ nm$^3$ with $Q=1.2$, connected to a second film of $2500 \times 3072 \times 2$ nm$^3$ with $Q=0$. A SW beam with $\theta=25^\circ$, $k_0 = 63$ rad/$\mu$m, and $f_0 = 1.5$ GHz is excited.

\subsubsection{Trireflection of spin waves in a 20-nm-thick CoFeB film}

For tri-reflection (Figure~\ref{fig:optics_thick}(b-c)), a system with dimensions $7620 \times 3072 \times 20$ nm$^3$ and $Q=0.6$ is connected to a second film of $1250 \times 3072 \times 20$ nm$^3$ with $Q=0$. A SW beam with $\theta = 65.5^\circ$, $k_0 = 130$ rad/$\mu$m, and $f_0 = 9$ GHz is excited. Figure 3(b-c) shows the results in the $(k_x, k_y)$-space and in the real space, respectively.

\subsubsection{Trirefraction of spin waves at a groove}

For tri-refraction at a groove (Figure \ref{fig:triRefraction} (a-c)), we simulate a system with dimensions $12800 \times 15360 \times 20$~nm$^3$ with $Q=0.6$. In the middle of the $x$-dimension a 10~nm thick and 20~nm wide groove is introduced, dividing the system into two halves. A SW beam with $\theta = 65.5^\circ$, $k_0 = 130$ rad/$\mu$m, and $f_0 = 9$ GHz is excited in the left part of the system (negative $x$ values). Fig. \ref{fig:triRefraction} (b), and (d-e) show the results in the real space and the ($k_x$, $k_y$)-space, respectively. 

\subsection{Torque components}
As described in the main part of the manuscript, equation~\ref{eq:linLLE_v2} can be expressed in terms of torques as:

\vspace{-.5cm}
\begin{equation}
    [\partial_t m_x, \partial_t m_z] =\boldsymbol{\tau}^\mathrm{eff} = [\tau^\mathrm{eff}_x, \tau^\mathrm{eff}_z],
\end{equation}

where the effective torque $\boldsymbol{\tau}^\mathrm{eff} $ consists of multiple contributions:  
$\boldsymbol{\tau}^\mathrm{eff} = \boldsymbol{\tau}^0 + \boldsymbol{\tau}^\mathrm{ex} + \boldsymbol{\tau}^\mathrm{d} + \boldsymbol{\tau}^\mathrm{PMA}$.
These terms correspond to torques originating from the external (Zeeman) field, exchange interactions, dipolar interactions, and PMA, respectively.  

For a spin wave propagating perpendicularly to the bias magnetic field (DE configuration), the individual torque components can be written as:  

\vspace{-.5cm}
\begin{align}
    \boldsymbol{\tau}^0 & = |\gamma|\mu_0 M_\mathrm{s}[(H_0/M_\mathrm{s})m_z,~ -(H_0/M_\mathrm{s})m_x]\\
    \boldsymbol{\tau}^\mathrm{d} &= |\gamma|\mu_0 M_\mathrm{s}[(1-\xi(kd)) m_z,~-\xi(kd) m_x  ] \\
    \boldsymbol{\tau}^\mathrm{ex} &= |\gamma|\mu_0 M_\mathrm{s}l_\mathrm{ex}^2 k^2 [m_z,~-m_x ]\\
    \boldsymbol{\tau}^\mathrm{PMA} &= |\gamma|\mu_0 M_\mathrm{s}[-Qm_z,~ 0].
\end{align}

\subsection{Dispersion relation}
Assuming a harmonic time dependence for $m_x$ and $m_z$ ($m_x, m_z \propto e^{-i \omega t}$), the dispersion relation of SWs can also be derived from equation~\ref{eq:linLLE_v2}.
\begin{equation}
    \omega = \sqrt{\omega_x \omega_z},
    \label{eq:dispersion}
\end{equation}
\vspace{-0.5cm}
where

\begin{align}
    \omega_x &= |\gamma|\mu_0 M_\mathrm{s} \left( H/M_\mathrm{s} + l_\mathrm{ex}^2 k^2 + h_{z}^{\mathrm{d}} - Q \right) \label{eq:wx}\\
    \omega_z &= |\gamma|\mu_0 M_\mathrm{s} \left( H/M_\mathrm{s} + l_\mathrm{ex}^2 k^2 + h_{x}^{\mathrm{d}} \right).\label{eq:wz}
\end{align} 

\vspace{0cm}

One may notice that $\omega_x$ and $\omega_z$ are proportional to the $x$- and $z$-components of the effective torque acting on magnetization.

\textbf{Supplementary Information} 
Derivation of the time dependencies of $m_x(t)$ and $m_z(t)$ for $\omega_x <0$, analysis of the $x$- and $z$- torque components in comparison to magnetization components across the film thickness, and a larger version of Fig~\ref{fig:optics_thick}(b).

\textbf{Movie S1.} An animated version of Figure~\ref{fig2:elipsy}(e-j) showing the precession of magnetic moments along the precession orbits.

\textbf{Movie S2.} An animation showing how the cross-sections (IFC) of the cowboy-hat-shaped dispersion profile from Figure~\ref{fig:optics_thick}(a) change with frequency.

\subsection*{Acknowledgement} 
The research leading to these results has received funding from the National Science Centre of Poland, project no. 2019/35/D/ST3/03729, and the Regional Council of Brittany (France) and ENIB, project AMOSPIN.
The authors thank M. Krawczyk and K. Szulc for valuable discussions.

\bibliography{literature}

\providecommand{\latin}[1]{#1}
\makeatletter
\providecommand{\doi}
  {\begingroup\let\do\@makeother\dospecials
  \catcode`\{=1 \catcode`\}=2 \doi@aux}
\providecommand{\doi@aux}[1]{\endgroup\texttt{#1}}
\makeatother
\providecommand*\mcitethebibliography{\thebibliography}
\csname @ifundefined\endcsname{endmcitethebibliography}  {\let\endmcitethebibliography\endthebibliography}{}
\begin{mcitethebibliography}{80}
\providecommand*\natexlab[1]{#1}
\providecommand*\mciteSetBstSublistMode[1]{}
\providecommand*\mciteSetBstMaxWidthForm[2]{}
\providecommand*\mciteBstWouldAddEndPuncttrue
  {\def\EndOfBibitem{\unskip.}}
\providecommand*\mciteBstWouldAddEndPunctfalse
  {\let\EndOfBibitem\relax}
\providecommand*\mciteSetBstMidEndSepPunct[3]{}
\providecommand*\mciteSetBstSublistLabelBeginEnd[3]{}
\providecommand*\EndOfBibitem{}
\mciteSetBstSublistMode{f}
\mciteSetBstMaxWidthForm{subitem}{(\alph{mcitesubitemcount})}
\mciteSetBstSublistLabelBeginEnd
  {\mcitemaxwidthsubitemform\space}
  {\relax}
  {\relax}

\bibitem[Rashed(1993)]{rashed1993}
Rashed,~R. \emph{G{\'e}om{\'e}trie et dioptrique au Xe si{\`e}cle: Ibn Sahl, al-Quhi et Ibn al-Haytham}; Les Belles Lettres: Paris, 1993\relax
\mciteBstWouldAddEndPuncttrue
\mciteSetBstMidEndSepPunct{\mcitedefaultmidpunct}
{\mcitedefaultendpunct}{\mcitedefaultseppunct}\relax
\EndOfBibitem
\bibitem[Descartes(1637)]{descartes1637}
Descartes,~R. \emph{Discours de la méthode pour bien conduire sa raison, et chercher la vérité dans les sciences}; Jan Maire: Leiden, Netherlands, 1637\relax
\mciteBstWouldAddEndPuncttrue
\mciteSetBstMidEndSepPunct{\mcitedefaultmidpunct}
{\mcitedefaultendpunct}{\mcitedefaultseppunct}\relax
\EndOfBibitem
\bibitem[Maxwell(1865)]{maxwell1865}
Maxwell,~J.~C. A dynamical theory of the electromagnetic field. \emph{Philosophical Transactions of the Royal Society of London} \textbf{1865}, \emph{155}, 459--512\relax
\mciteBstWouldAddEndPuncttrue
\mciteSetBstMidEndSepPunct{\mcitedefaultmidpunct}
{\mcitedefaultendpunct}{\mcitedefaultseppunct}\relax
\EndOfBibitem
\bibitem[Hecht(2016)]{hecht2016}
Hecht,~E. \emph{Optics}, 5th ed.; Pearson, 2016\relax
\mciteBstWouldAddEndPuncttrue
\mciteSetBstMidEndSepPunct{\mcitedefaultmidpunct}
{\mcitedefaultendpunct}{\mcitedefaultseppunct}\relax
\EndOfBibitem
\bibitem[Gruszecki \latin{et~al.}(2014)Gruszecki, Romero-Vivas, Dadoenkova, Dadoenkova, Lyubchanskii, and Krawczyk]{gruszecki2014}
Gruszecki,~P.; Romero-Vivas,~J.; Dadoenkova,~Y.~S.; Dadoenkova,~N.; Lyubchanskii,~I.; Krawczyk,~M. Goos-{H}{\"a}nchen effect and bending of spin wave beams in thin magnetic films. \emph{Applied Physics Letters} \textbf{2014}, \emph{105}\relax
\mciteBstWouldAddEndPuncttrue
\mciteSetBstMidEndSepPunct{\mcitedefaultmidpunct}
{\mcitedefaultendpunct}{\mcitedefaultseppunct}\relax
\EndOfBibitem
\bibitem[Stigloher \latin{et~al.}(2016)Stigloher, Decker, K{\"o}rner, Tanabe, Moriyama, Taniguchi, Hata, Madami, Gubbiotti, Kobayashi, \latin{et~al.} others]{stigloher2016}
Stigloher,~J.; Decker,~M.; K{\"o}rner,~H.~S.; Tanabe,~K.; Moriyama,~T.; Taniguchi,~T.; Hata,~H.; Madami,~M.; Gubbiotti,~G.; Kobayashi,~K.; others Snell’s law for spin waves. \emph{Physical Review Letters} \textbf{2016}, \emph{117}, 037204\relax
\mciteBstWouldAddEndPuncttrue
\mciteSetBstMidEndSepPunct{\mcitedefaultmidpunct}
{\mcitedefaultendpunct}{\mcitedefaultseppunct}\relax
\EndOfBibitem
\bibitem[Yu \latin{et~al.}(2016)Yu, Lan, Wu, and Xiao]{yu2016}
Yu,~W.; Lan,~J.; Wu,~R.; Xiao,~J. Magnetic {S}nell's law and spin-wave fiber with {D}zyaloshinskii-{M}oriya interaction. \emph{Physical Review B} \textbf{2016}, \emph{94}, 140410\relax
\mciteBstWouldAddEndPuncttrue
\mciteSetBstMidEndSepPunct{\mcitedefaultmidpunct}
{\mcitedefaultendpunct}{\mcitedefaultseppunct}\relax
\EndOfBibitem
\bibitem[Mulkers \latin{et~al.}(2018)Mulkers, Van~Waeyenberge, and Milo{\v{s}}evi{\'c}]{mulkers2018}
Mulkers,~J.; Van~Waeyenberge,~B.; Milo{\v{s}}evi{\'c},~M.~V. Tunable {S}nell's law for spin waves in heterochiral magnetic films. \emph{Physical Review B} \textbf{2018}, \emph{97}, 104422\relax
\mciteBstWouldAddEndPuncttrue
\mciteSetBstMidEndSepPunct{\mcitedefaultmidpunct}
{\mcitedefaultendpunct}{\mcitedefaultseppunct}\relax
\EndOfBibitem
\bibitem[Gruszecki and Krawczyk(2018)Gruszecki, and Krawczyk]{gruszecki2018}
Gruszecki,~P.; Krawczyk,~M. Spin-wave beam propagation in ferromagnetic thin films with graded refractive index: Mirage effect and prospective applications. \emph{Physical Review B} \textbf{2018}, \emph{97}, 094424\relax
\mciteBstWouldAddEndPuncttrue
\mciteSetBstMidEndSepPunct{\mcitedefaultmidpunct}
{\mcitedefaultendpunct}{\mcitedefaultseppunct}\relax
\EndOfBibitem
\bibitem[Zhu \latin{et~al.}(2022)Zhu, Qin, Flaj{\v{s}}man, Taniyama, and Van~Dijken]{zhu2022}
Zhu,~W.; Qin,~H.; Flaj{\v{s}}man,~L.; Taniyama,~T.; Van~Dijken,~S. Zero-field routing of spin waves in a multiferroic heterostructure. \emph{Applied Physics Letters} \textbf{2022}, \emph{120}, 112407\relax
\mciteBstWouldAddEndPuncttrue
\mciteSetBstMidEndSepPunct{\mcitedefaultmidpunct}
{\mcitedefaultendpunct}{\mcitedefaultseppunct}\relax
\EndOfBibitem
\bibitem[Yu and Capasso(2014)Yu, and Capasso]{Yu2014}
Yu,~N.; Capasso,~F. Flat optics with designer metasurfaces. \emph{Nature Materials} \textbf{2014}, \emph{13}, 139–150\relax
\mciteBstWouldAddEndPuncttrue
\mciteSetBstMidEndSepPunct{\mcitedefaultmidpunct}
{\mcitedefaultendpunct}{\mcitedefaultseppunct}\relax
\EndOfBibitem
\bibitem[Ni \latin{et~al.}(2012)Ni, Emani, Kildishev, Boltasseva, and Shalaev]{Ni2012}
Ni,~X.; Emani,~N.~K.; Kildishev,~A.~V.; Boltasseva,~A.; Shalaev,~V.~M. Broadband Light Bending with Plasmonic Nanoantennas. \emph{Science} \textbf{2012}, \emph{335}, 427–427\relax
\mciteBstWouldAddEndPuncttrue
\mciteSetBstMidEndSepPunct{\mcitedefaultmidpunct}
{\mcitedefaultendpunct}{\mcitedefaultseppunct}\relax
\EndOfBibitem
\bibitem[Pfeiffer and Grbic(2013)Pfeiffer, and Grbic]{Pfeiffer2013}
Pfeiffer,~C.; Grbic,~A. Metamaterial Huygens’ Surfaces: Tailoring Wave Fronts with Reflectionless Sheets. \emph{Physical Review Letters} \textbf{2013}, \emph{110}\relax
\mciteBstWouldAddEndPuncttrue
\mciteSetBstMidEndSepPunct{\mcitedefaultmidpunct}
{\mcitedefaultendpunct}{\mcitedefaultseppunct}\relax
\EndOfBibitem
\bibitem[Sun \latin{et~al.}(2012)Sun, Yang, Wang, Juan, Chen, Liao, He, Xiao, Kung, Guo, Zhou, and Tsai]{Sun2012}
Sun,~S.; Yang,~K.-Y.; Wang,~C.-M.; Juan,~T.-K.; Chen,~W.~T.; Liao,~C.~Y.; He,~Q.; Xiao,~S.; Kung,~W.-T.; Guo,~G.-Y.; Zhou,~L.; Tsai,~D.~P. High-Efficiency Broadband Anomalous Reflection by Gradient Meta-Surfaces. \emph{Nano Letters} \textbf{2012}, \emph{12}, 6223–6229\relax
\mciteBstWouldAddEndPuncttrue
\mciteSetBstMidEndSepPunct{\mcitedefaultmidpunct}
{\mcitedefaultendpunct}{\mcitedefaultseppunct}\relax
\EndOfBibitem
\bibitem[Poddubny \latin{et~al.}(2013)Poddubny, Iorsh, Belov, and Kivshar]{Poddubny2013}
Poddubny,~A.; Iorsh,~I.; Belov,~P.; Kivshar,~Y. Hyperbolic metamaterials. \emph{Nature Photonics} \textbf{2013}, \emph{7}, 948–957\relax
\mciteBstWouldAddEndPuncttrue
\mciteSetBstMidEndSepPunct{\mcitedefaultmidpunct}
{\mcitedefaultendpunct}{\mcitedefaultseppunct}\relax
\EndOfBibitem
\bibitem[Mac\^edo and Dumelow(2014)Mac\^edo, and Dumelow]{Macdo2014}
Mac\^edo,~R.; Dumelow,~T. Tunable all-angle negative refraction using antiferromagnets. \emph{Physical Review B} \textbf{2014}, \emph{89}\relax
\mciteBstWouldAddEndPuncttrue
\mciteSetBstMidEndSepPunct{\mcitedefaultmidpunct}
{\mcitedefaultendpunct}{\mcitedefaultseppunct}\relax
\EndOfBibitem
\bibitem[Zhang \latin{et~al.}(2022)Zhang, Zheng, Chen, and Qiu]{Zhang2022}
Zhang,~T.; Zheng,~C.; Chen,~Z.~N.; Qiu,~C.-W. Negative Reflection and Negative Refraction in Biaxial van der {W}aals Materials. \emph{Nano Letters} \textbf{2022}, \emph{22}, 5607–5614\relax
\mciteBstWouldAddEndPuncttrue
\mciteSetBstMidEndSepPunct{\mcitedefaultmidpunct}
{\mcitedefaultendpunct}{\mcitedefaultseppunct}\relax
\EndOfBibitem
\bibitem[Gérardin \latin{et~al.}(2019)Gérardin, Laurent, Legrand, Prada, and Aubry]{Grardin2019}
Gérardin,~B.; Laurent,~J.; Legrand,~F.; Prada,~C.; Aubry,~A. Negative reflection of elastic guided waves in chaotic and random scattering media. \emph{Scientific Reports} \textbf{2019}, \emph{9}\relax
\mciteBstWouldAddEndPuncttrue
\mciteSetBstMidEndSepPunct{\mcitedefaultmidpunct}
{\mcitedefaultendpunct}{\mcitedefaultseppunct}\relax
\EndOfBibitem
\bibitem[Bang \latin{et~al.}(2019)Bang, So, and Rho]{Bang2019}
Bang,~S.; So,~S.; Rho,~J. Realization of broadband negative refraction in visible range using vertically stacked hyperbolic metamaterials. \emph{Scientific Reports} \textbf{2019}, \emph{9}\relax
\mciteBstWouldAddEndPuncttrue
\mciteSetBstMidEndSepPunct{\mcitedefaultmidpunct}
{\mcitedefaultendpunct}{\mcitedefaultseppunct}\relax
\EndOfBibitem
\bibitem[Jaksic \latin{et~al.}(2006)Jaksic, Dalarsson, and Maksimovic]{jaksic2006}
Jaksic,~Z.; Dalarsson,~N.; Maksimovic,~M. Negative Refractive Index Metamaterials: Principles and Applications. \emph{Microwave Review} \textbf{2006}, \emph{12}\relax
\mciteBstWouldAddEndPuncttrue
\mciteSetBstMidEndSepPunct{\mcitedefaultmidpunct}
{\mcitedefaultendpunct}{\mcitedefaultseppunct}\relax
\EndOfBibitem
\bibitem[Zhang \latin{et~al.}(2009)Zhang, Park, Li, Lu, Zhang, and Zhang]{Zhang2009}
Zhang,~S.; Park,~Y.-S.; Li,~J.; Lu,~X.; Zhang,~W.; Zhang,~X. Negative Refractive Index in Chiral Metamaterials. \emph{Physical Review Letters} \textbf{2009}, \emph{102}\relax
\mciteBstWouldAddEndPuncttrue
\mciteSetBstMidEndSepPunct{\mcitedefaultmidpunct}
{\mcitedefaultendpunct}{\mcitedefaultseppunct}\relax
\EndOfBibitem
\bibitem[Kim \latin{et~al.}(2008)Kim, Choi, Lee, Han, Jung, and Choi]{kim2008}
Kim,~S.-K.; Choi,~S.; Lee,~K.-S.; Han,~D.-S.; Jung,~D.-E.; Choi,~Y.-S. Negative refraction of dipole-exchange spin waves through a magnetic twin interface in restricted geometry. \emph{Applied Physics Letters} \textbf{2008}, \emph{92}, 212501\relax
\mciteBstWouldAddEndPuncttrue
\mciteSetBstMidEndSepPunct{\mcitedefaultmidpunct}
{\mcitedefaultendpunct}{\mcitedefaultseppunct}\relax
\EndOfBibitem
\bibitem[Hioki \latin{et~al.}(2020)Hioki, Hashimoto, and Saitoh]{hioki2020}
Hioki,~T.; Hashimoto,~Y.; Saitoh,~E. Bi-reflection of spin waves. \emph{Communications Physics} \textbf{2020}, \emph{3}, 188\relax
\mciteBstWouldAddEndPuncttrue
\mciteSetBstMidEndSepPunct{\mcitedefaultmidpunct}
{\mcitedefaultendpunct}{\mcitedefaultseppunct}\relax
\EndOfBibitem
\bibitem[Fleury \latin{et~al.}(2014)Fleury, Sounas, and Alu]{fleury2014}
Fleury,~R.; Sounas,~D.~L.; Alu,~A. Negative refraction and planar focusing based on parity-time symmetric metasurfaces. \emph{Physical Review Letters} \textbf{2014}, \emph{113}, 023903\relax
\mciteBstWouldAddEndPuncttrue
\mciteSetBstMidEndSepPunct{\mcitedefaultmidpunct}
{\mcitedefaultendpunct}{\mcitedefaultseppunct}\relax
\EndOfBibitem
\bibitem[Liu \latin{et~al.}(2017)Liu, Ren, Zhao, Xu, Feng, Zhao, and Jiang]{liu2017}
Liu,~B.; Ren,~B.; Zhao,~J.; Xu,~X.; Feng,~Y.; Zhao,~W.; Jiang,~Y. Experimental realization of all-angle negative refraction in acoustic gradient metasurface. \emph{Applied Physics Letters} \textbf{2017}, \emph{111}\relax
\mciteBstWouldAddEndPuncttrue
\mciteSetBstMidEndSepPunct{\mcitedefaultmidpunct}
{\mcitedefaultendpunct}{\mcitedefaultseppunct}\relax
\EndOfBibitem
\bibitem[Mieszczak \latin{et~al.}(2020)Mieszczak, Busel, Gruszecki, Kuchko, K{\l}os, and Krawczyk]{mieszczak2020}
Mieszczak,~S.; Busel,~O.; Gruszecki,~P.; Kuchko,~A.~N.; K{\l}os,~J.~W.; Krawczyk,~M. Anomalous refraction of spin waves as a way to guide signals in curved magnonic multimode waveguides. \emph{Physical Review Applied} \textbf{2020}, \emph{13}, 054038\relax
\mciteBstWouldAddEndPuncttrue
\mciteSetBstMidEndSepPunct{\mcitedefaultmidpunct}
{\mcitedefaultendpunct}{\mcitedefaultseppunct}\relax
\EndOfBibitem
\bibitem[Zelent \latin{et~al.}(2019)Zelent, Mailyan, Vashistha, Gruszecki, Gorobets, Gorobets, and Krawczyk]{zelent2019}
Zelent,~M.; Mailyan,~M.; Vashistha,~V.; Gruszecki,~P.; Gorobets,~O.; Gorobets,~Y.; Krawczyk,~M. Spin wave collimation using a flat metasurface. \emph{Nanoscale} \textbf{2019}, \emph{11}, 9743--9748\relax
\mciteBstWouldAddEndPuncttrue
\mciteSetBstMidEndSepPunct{\mcitedefaultmidpunct}
{\mcitedefaultendpunct}{\mcitedefaultseppunct}\relax
\EndOfBibitem
\bibitem[Wang \latin{et~al.}(2018)Wang, Ding, Zheng, An, Zhai, and Zhang]{Wang2018}
Wang,~X.; Ding,~J.; Zheng,~B.; An,~S.; Zhai,~G.; Zhang,~H. Simultaneous Realization of Anomalous Reflection and Transmission at Two Frequencies using Bi-functional Metasurfaces. \emph{Scientific Reports} \textbf{2018}, \emph{8}, 1876\relax
\mciteBstWouldAddEndPuncttrue
\mciteSetBstMidEndSepPunct{\mcitedefaultmidpunct}
{\mcitedefaultendpunct}{\mcitedefaultseppunct}\relax
\EndOfBibitem
\bibitem[Meirbekova \latin{et~al.}(2023)Meirbekova, Morini, Brun, and Carta]{Meirbekova2023}
Meirbekova,~B.; Morini,~L.; Brun,~M.; Carta,~G. The strange case of negative reflection. \emph{Applied Physics Letters} \textbf{2023}, \emph{123}, 031704\relax
\mciteBstWouldAddEndPuncttrue
\mciteSetBstMidEndSepPunct{\mcitedefaultmidpunct}
{\mcitedefaultendpunct}{\mcitedefaultseppunct}\relax
\EndOfBibitem
\bibitem[Liu \latin{et~al.}(2018)Liu, Jun~Cui, Noor, Tao, Chi~Zhang, Dong~Bai, Yang, and Yang~Zhou]{liu2018}
Liu,~S.; Jun~Cui,~T.; Noor,~A.; Tao,~Z.; Chi~Zhang,~H.; Dong~Bai,~G.; Yang,~Y.; Yang~Zhou,~X. Negative reflection and negative surface wave conversion from obliquely incident electromagnetic waves. \emph{Light: Science \& Applications} \textbf{2018}, \emph{7}, 18008--18008\relax
\mciteBstWouldAddEndPuncttrue
\mciteSetBstMidEndSepPunct{\mcitedefaultmidpunct}
{\mcitedefaultendpunct}{\mcitedefaultseppunct}\relax
\EndOfBibitem
\bibitem[{\'A}lvarez-P{\'e}rez \latin{et~al.}(2022){\'A}lvarez-P{\'e}rez, Duan, Taboada-Guti{\'e}rrez, Ou, Nikulina, Liu, Edgar, Bao, Giannini, Hillenbrand, \latin{et~al.} others]{alvarez2022}
{\'A}lvarez-P{\'e}rez,~G.; Duan,~J.; Taboada-Guti{\'e}rrez,~J.; Ou,~Q.; Nikulina,~E.; Liu,~S.; Edgar,~J.~H.; Bao,~Q.; Giannini,~V.; Hillenbrand,~R.; others Negative reflection of nanoscale-confined polaritons in a low-loss natural medium. \emph{Science Advances} \textbf{2022}, \emph{8}, eabp8486\relax
\mciteBstWouldAddEndPuncttrue
\mciteSetBstMidEndSepPunct{\mcitedefaultmidpunct}
{\mcitedefaultendpunct}{\mcitedefaultseppunct}\relax
\EndOfBibitem
\bibitem[Dadoenkova \latin{et~al.}(2022)Dadoenkova, Glukhov, Moiseev, and Bentivegna]{dadoenkova2022}
Dadoenkova,~Y.~S.; Glukhov,~I.~A.; Moiseev,~S.~G.; Bentivegna,~F.~F. Non-specular reflection of a narrow spatially phase-modulated Gaussian beam. \emph{JOSA A} \textbf{2022}, \emph{39}, 2073--2082\relax
\mciteBstWouldAddEndPuncttrue
\mciteSetBstMidEndSepPunct{\mcitedefaultmidpunct}
{\mcitedefaultendpunct}{\mcitedefaultseppunct}\relax
\EndOfBibitem
\bibitem[Stancil and Prabhakar(2009)Stancil, and Prabhakar]{stancil}
Stancil,~D.~D.; Prabhakar,~A. \emph{Spin Waves: Theory and Applications}; Springer: Boston, MA, 2009\relax
\mciteBstWouldAddEndPuncttrue
\mciteSetBstMidEndSepPunct{\mcitedefaultmidpunct}
{\mcitedefaultendpunct}{\mcitedefaultseppunct}\relax
\EndOfBibitem
\bibitem[Hubert and Sch{\"a}fer(2008)Hubert, and Sch{\"a}fer]{hubert2008}
Hubert,~A.; Sch{\"a}fer,~R. \emph{Magnetic domains: the analysis of magnetic microstructures}; Springer Science \& Business Media, 2008\relax
\mciteBstWouldAddEndPuncttrue
\mciteSetBstMidEndSepPunct{\mcitedefaultmidpunct}
{\mcitedefaultendpunct}{\mcitedefaultseppunct}\relax
\EndOfBibitem
\bibitem[Kumar~Mishra \latin{et~al.}(2024)Kumar~Mishra, Prasanth~Perumal, and Mohanty]{KumarMishra2024}
Kumar~Mishra,~S.; Prasanth~Perumal,~H.; Mohanty,~J. Engineering perpendicular magnetic anisotropy and {D}zyaloshinskii–{M}oriya interaction in {G}d–{F}e thin films for spintronics applications. \emph{Journal of Applied Physics} \textbf{2024}, \emph{136}, 243901\relax
\mciteBstWouldAddEndPuncttrue
\mciteSetBstMidEndSepPunct{\mcitedefaultmidpunct}
{\mcitedefaultendpunct}{\mcitedefaultseppunct}\relax
\EndOfBibitem
\bibitem[Fallarino \latin{et~al.}(2019)Fallarino, Oelschl{\"a}gel, Arregi, Bashkatov, Samad, B{\"o}hm, Chesnel, and Hellwig]{fallarino2019}
Fallarino,~L.; Oelschl{\"a}gel,~A.; Arregi,~J.; Bashkatov,~A.; Samad,~F.; B{\"o}hm,~B.; Chesnel,~K.; Hellwig,~O. Control of domain structure and magnetization reversal in thick {C}o/{P}t multilayers. \emph{Physical Review B} \textbf{2019}, \emph{99}, 024431\relax
\mciteBstWouldAddEndPuncttrue
\mciteSetBstMidEndSepPunct{\mcitedefaultmidpunct}
{\mcitedefaultendpunct}{\mcitedefaultseppunct}\relax
\EndOfBibitem
\bibitem[Vukadinovic \latin{et~al.}(2000)Vukadinovic, Labrune, Pain, \latin{et~al.} others]{vukadinovic2000}
Vukadinovic,~N.; Labrune,~M.; Pain,~D.; others Magnetic excitations in a weak-stripe-domain structure: A {2D} dynamic micromagnetic approach. \emph{Physical Review Letters} \textbf{2000}, \emph{85}, 2817\relax
\mciteBstWouldAddEndPuncttrue
\mciteSetBstMidEndSepPunct{\mcitedefaultmidpunct}
{\mcitedefaultendpunct}{\mcitedefaultseppunct}\relax
\EndOfBibitem
\bibitem[Mondal \latin{et~al.}(2019)Mondal, Talapatra, Arout~Chelvane, Mohanty, and Barman]{Sucheta2019}
Mondal,~S.; Talapatra,~A.; Arout~Chelvane,~J.; Mohanty,~J.~R.; Barman,~A. Role of magnetic anisotropy in the ultrafast magnetization dynamics of {G}d-{F}e thin films of different thicknesses. \emph{Phys. Rev. B} \textbf{2019}, \emph{100}, 054436\relax
\mciteBstWouldAddEndPuncttrue
\mciteSetBstMidEndSepPunct{\mcitedefaultmidpunct}
{\mcitedefaultendpunct}{\mcitedefaultseppunct}\relax
\EndOfBibitem
\bibitem[Grassi \latin{et~al.}(2022)Grassi, Geilen, Oukaci, Henry, Lacour, Stoeffler, Hehn, Pirro, and Bailleul]{grassi2022}
Grassi,~M.; Geilen,~M.; Oukaci,~K.~A.; Henry,~Y.; Lacour,~D.; Stoeffler,~D.; Hehn,~M.; Pirro,~P.; Bailleul,~M. Higgs and {G}oldstone spin-wave modes in striped magnetic texture. \emph{Phys. Rev. B} \textbf{2022}, \emph{105}, 094444\relax
\mciteBstWouldAddEndPuncttrue
\mciteSetBstMidEndSepPunct{\mcitedefaultmidpunct}
{\mcitedefaultendpunct}{\mcitedefaultseppunct}\relax
\EndOfBibitem
\bibitem[Alvarez-Prado(2021)]{alvarez2021}
Alvarez-Prado,~L.~M. Control of dynamics in weak {PMA} magnets. \emph{Magnetochemistry} \textbf{2021}, \emph{7}, 43\relax
\mciteBstWouldAddEndPuncttrue
\mciteSetBstMidEndSepPunct{\mcitedefaultmidpunct}
{\mcitedefaultendpunct}{\mcitedefaultseppunct}\relax
\EndOfBibitem
\bibitem[Janardhanan \latin{et~al.}(2024)Janardhanan, Krawczyk, and Trzaskowska]{Janardhanan2024}
Janardhanan,~S.; Krawczyk,~M.; Trzaskowska,~A. In \emph{Nanomagnets as Dynamical Systems: Physics and Applications}; Bandyopadhyay,~S., Barman,~A., Eds.; Springer Nature Switzerland: Cham, 2024; pp 33--69\relax
\mciteBstWouldAddEndPuncttrue
\mciteSetBstMidEndSepPunct{\mcitedefaultmidpunct}
{\mcitedefaultendpunct}{\mcitedefaultseppunct}\relax
\EndOfBibitem
\bibitem[Banerjee \latin{et~al.}(2017)Banerjee, Gruszecki, Klos, Hellwig, Krawczyk, and Barman]{banerjee2017}
Banerjee,~C.; Gruszecki,~P.; Klos,~J.~W.; Hellwig,~O.; Krawczyk,~M.; Barman,~A. Magnonic band structure in a {C}o/{P}d stripe domain system investigated by {B}rillouin light scattering and micromagnetic simulations. \emph{Physical Review B} \textbf{2017}, \emph{96}, 024421\relax
\mciteBstWouldAddEndPuncttrue
\mciteSetBstMidEndSepPunct{\mcitedefaultmidpunct}
{\mcitedefaultendpunct}{\mcitedefaultseppunct}\relax
\EndOfBibitem
\bibitem[Kisielewski \latin{et~al.}(2023)Kisielewski, Gruszecki, Krawczyk, Zablotskii, and Maziewski]{kisielewski2023}
Kisielewski,~J.; Gruszecki,~P.; Krawczyk,~M.; Zablotskii,~V.; Maziewski,~A. Between waves and patterns: Spin wave freezing in films with {D}zyaloshinskii-{M}oriya interaction. \emph{Physical Review B} \textbf{2023}, \emph{107}, 134416\relax
\mciteBstWouldAddEndPuncttrue
\mciteSetBstMidEndSepPunct{\mcitedefaultmidpunct}
{\mcitedefaultendpunct}{\mcitedefaultseppunct}\relax
\EndOfBibitem
\bibitem[Yoshizawa \latin{et~al.}(1985)Yoshizawa, M.~Shapiro, and Komatsubara]{yoshizawa1985}
Yoshizawa,~H.; M.~Shapiro,~S.; Komatsubara,~T. Softening of the Spin Waves in MnP and Its Relation to the Lifshitz Point. \emph{Journal of the Physical Society of Japan} \textbf{1985}, \emph{54}, 3084--3090\relax
\mciteBstWouldAddEndPuncttrue
\mciteSetBstMidEndSepPunct{\mcitedefaultmidpunct}
{\mcitedefaultendpunct}{\mcitedefaultseppunct}\relax
\EndOfBibitem
\bibitem[Steiner \latin{et~al.}(1996)Steiner, Kakurai, Dorner, Pynn, and Akimitsu]{Steiner1996}
Steiner,~M.; Kakurai,~K.; Dorner,~B.; Pynn,~R.; Akimitsu,~J. The change of the symmetry of the soft-mode spin wave at the spin - flop transition as observed in using inelastic polarized neutron scattering. \emph{Journal of Physics: Condensed Matter} \textbf{1996}, \emph{8}, 5905\relax
\mciteBstWouldAddEndPuncttrue
\mciteSetBstMidEndSepPunct{\mcitedefaultmidpunct}
{\mcitedefaultendpunct}{\mcitedefaultseppunct}\relax
\EndOfBibitem
\bibitem[Sun and Lin(2006)Sun, and Lin]{Sun2006}
Sun,~S.-J.; Lin,~H.-H. Softening of spin-wave stiffness near the ferromagnetic phase transition in diluted magnetic semiconductors. \emph{The European Physical Journal B - Condensed Matter and Complex Systems} \textbf{2006}, \emph{49}, 403–406\relax
\mciteBstWouldAddEndPuncttrue
\mciteSetBstMidEndSepPunct{\mcitedefaultmidpunct}
{\mcitedefaultendpunct}{\mcitedefaultseppunct}\relax
\EndOfBibitem
\bibitem[Bailleul \latin{et~al.}(2003)Bailleul, Olligs, and Fermon]{bailleul2003}
Bailleul,~M.; Olligs,~D.; Fermon,~C. Micromagnetic phase transitions and spin wave excitations in a ferromagnetic stripe. \emph{Physical review letters} \textbf{2003}, \emph{91}, 137204\relax
\mciteBstWouldAddEndPuncttrue
\mciteSetBstMidEndSepPunct{\mcitedefaultmidpunct}
{\mcitedefaultendpunct}{\mcitedefaultseppunct}\relax
\EndOfBibitem
\bibitem[Leaf \latin{et~al.}(2006)Leaf, Kaper, Yan, Novosad, Vavassori, Camley, and Grimsditch]{Leaf2006}
Leaf,~G.; Kaper,~H.; Yan,~M.; Novosad,~V.; Vavassori,~P.; Camley,~R.~E.; Grimsditch,~M. Dynamic Origin of Stripe Domains. \emph{Phys. Rev. Lett.} \textbf{2006}, \emph{96}, 017201\relax
\mciteBstWouldAddEndPuncttrue
\mciteSetBstMidEndSepPunct{\mcitedefaultmidpunct}
{\mcitedefaultendpunct}{\mcitedefaultseppunct}\relax
\EndOfBibitem
\bibitem[Kalinikos and Slavin(1986)Kalinikos, and Slavin]{kalinikos1986}
Kalinikos,~B.; Slavin,~A. Theory of dipole-exchange spin wave spectrum for ferromagnetic films with mixed exchange boundary conditions. \emph{Journal of Physics C: Solid State Physics} \textbf{1986}, \emph{19}, 7013\relax
\mciteBstWouldAddEndPuncttrue
\mciteSetBstMidEndSepPunct{\mcitedefaultmidpunct}
{\mcitedefaultendpunct}{\mcitedefaultseppunct}\relax
\EndOfBibitem
\bibitem[Gruszecki \latin{et~al.}(2015)Gruszecki, Dadoenkova, Dadoenkova, Lyubchanskii, Romero-Vivas, Guslienko, and Krawczyk]{gruszecki2015}
Gruszecki,~P.; Dadoenkova,~Y.~S.; Dadoenkova,~N.; Lyubchanskii,~I.; Romero-Vivas,~J.; Guslienko,~K.; Krawczyk,~M. Influence of magnetic surface anisotropy on spin wave reflection from the edge of ferromagnetic film. \emph{Physical Review B} \textbf{2015}, \emph{92}, 054427\relax
\mciteBstWouldAddEndPuncttrue
\mciteSetBstMidEndSepPunct{\mcitedefaultmidpunct}
{\mcitedefaultendpunct}{\mcitedefaultseppunct}\relax
\EndOfBibitem
\bibitem[Gurevich and Melkov(1996)Gurevich, and Melkov]{GurevichMelkov1996}
Gurevich,~A.~G.; Melkov,~G.~A. \emph{Magnetization Oscillations and Waves}; CRC Press: Boca Raton, 1996\relax
\mciteBstWouldAddEndPuncttrue
\mciteSetBstMidEndSepPunct{\mcitedefaultmidpunct}
{\mcitedefaultendpunct}{\mcitedefaultseppunct}\relax
\EndOfBibitem
\bibitem[Lifshitz and Pitaevskii(1980)Lifshitz, and Pitaevskii]{landau1980}
Lifshitz,~E.~M.; Pitaevskii,~L.~P. Statistical physics: theory of the condensed state. \textbf{1980}, \emph{9}\relax
\mciteBstWouldAddEndPuncttrue
\mciteSetBstMidEndSepPunct{\mcitedefaultmidpunct}
{\mcitedefaultendpunct}{\mcitedefaultseppunct}\relax
\EndOfBibitem
\bibitem[Vansteenkiste \latin{et~al.}(2014)Vansteenkiste, Leliaert, Dvornik, Helsen, Garcia-Sanchez, and Van~Waeyenberge]{Vansteenkiste2014}
Vansteenkiste,~A.; Leliaert,~J.; Dvornik,~M.; Helsen,~M.; Garcia-Sanchez,~F.; Van~Waeyenberge,~B. The design and verification of MuMax3. \emph{AIP advances} \textbf{2014}, \emph{4}, 107133\relax
\mciteBstWouldAddEndPuncttrue
\mciteSetBstMidEndSepPunct{\mcitedefaultmidpunct}
{\mcitedefaultendpunct}{\mcitedefaultseppunct}\relax
\EndOfBibitem
\bibitem[Demokritov \latin{et~al.}(2006)Demokritov, Demidov, Dzyapko, Melkov, Serga, Hillebrands, and Slavin]{demokritov2006}
Demokritov,~S.~O.; Demidov,~V.~E.; Dzyapko,~O.; Melkov,~G.~A.; Serga,~A.~A.; Hillebrands,~B.; Slavin,~A.~N. Bose--{E}instein condensation of quasi-equilibrium magnons at room temperature under pumping. \emph{Nature} \textbf{2006}, \emph{443}, 430--433\relax
\mciteBstWouldAddEndPuncttrue
\mciteSetBstMidEndSepPunct{\mcitedefaultmidpunct}
{\mcitedefaultendpunct}{\mcitedefaultseppunct}\relax
\EndOfBibitem
\bibitem[Eshbach and Damon(1960)Eshbach, and Damon]{Damon_Eshbach}
Eshbach,~J.~R.; Damon,~R.~W. Surface Magnetostatic Modes and Surface Spin Waves. \emph{Phys. Rev.} \textbf{1960}, \emph{118}, 1208--1210\relax
\mciteBstWouldAddEndPuncttrue
\mciteSetBstMidEndSepPunct{\mcitedefaultmidpunct}
{\mcitedefaultendpunct}{\mcitedefaultseppunct}\relax
\EndOfBibitem
\bibitem[Rych{\l}y \latin{et~al.}(2016)Rych{\l}y, K{\l}os, and Krawczyk]{rychly2016}
Rych{\l}y,~J.; K{\l}os,~J.~W.; Krawczyk,~M. Spin wave damping in periodic and quasiperiodic magnonic structures. \emph{Journal of Physics D: Applied Physics} \textbf{2016}, \emph{49}, 175001\relax
\mciteBstWouldAddEndPuncttrue
\mciteSetBstMidEndSepPunct{\mcitedefaultmidpunct}
{\mcitedefaultendpunct}{\mcitedefaultseppunct}\relax
\EndOfBibitem
\bibitem[Szulc \latin{et~al.}(2022)Szulc, Tacchi, Hierro-Rodríguez, Díaz, Gruszecki, Graczyk, Quirós, Markó, Martín, V{\'e}lez, Schmool, Carlotti, Krawczyk, and Álvarez Prado]{Szulc2022}
Szulc,~K.; Tacchi,~S.; Hierro-Rodríguez,~A.; Díaz,~J.; Gruszecki,~P.; Graczyk,~P.; Quirós,~C.; Markó,~D.; Martín,~J.~I.; V{\'e}lez,~M.; Schmool,~D.~S.; Carlotti,~G.; Krawczyk,~M.; Álvarez Prado,~L.~M. Reconfigurable Magnonic Crystals Based on Imprinted Magnetization Textures in Hard and Soft Dipolar-Coupled Bilayers. \emph{ACS Nano} \textbf{2022}, \emph{16}, 14168--14177, PMID: 36043881\relax
\mciteBstWouldAddEndPuncttrue
\mciteSetBstMidEndSepPunct{\mcitedefaultmidpunct}
{\mcitedefaultendpunct}{\mcitedefaultseppunct}\relax
\EndOfBibitem
\bibitem[Szulc \latin{et~al.}(2024)Szulc, Kharlan, Bondarenko, Tartakovskaya, and Krawczyk]{Szulc2024}
Szulc,~K.; Kharlan,~J.; Bondarenko,~P.; Tartakovskaya,~E.~V.; Krawczyk,~M. Impact of surface anisotropy on the spin-wave dynamics in a thin ferromagnetic film. \emph{Phys. Rev. B} \textbf{2024}, \emph{109}, 054430\relax
\mciteBstWouldAddEndPuncttrue
\mciteSetBstMidEndSepPunct{\mcitedefaultmidpunct}
{\mcitedefaultendpunct}{\mcitedefaultseppunct}\relax
\EndOfBibitem
\bibitem[Dhiman \latin{et~al.}(2024)Dhiman, Le{\'s}niewski, Gieniusz, Kisielewski, Mazalski, Kurant, Matczak, Stobiecki, Krawczyk, Lynnyk, \latin{et~al.} others]{dhiman2024}
Dhiman,~A.~K.; Le{\'s}niewski,~N.; Gieniusz,~R.; Kisielewski,~J.; Mazalski,~P.; Kurant,~Z.; Matczak,~M.; Stobiecki,~F.; Krawczyk,~M.; Lynnyk,~A.; others Reconfigurable magnonic crystals: Spin wave propagation in {P}t/{C}o multilayer in saturated and stripe domain phase. \emph{APL Materials} \textbf{2024}, \emph{12}\relax
\mciteBstWouldAddEndPuncttrue
\mciteSetBstMidEndSepPunct{\mcitedefaultmidpunct}
{\mcitedefaultendpunct}{\mcitedefaultseppunct}\relax
\EndOfBibitem
\bibitem[Salanskii and Erukhimov(1975)Salanskii, and Erukhimov]{salanskii1975}
Salanskii,~N.; Erukhimov,~M.~S. Physical Properties and Application of Magnetic Films. 1975\relax
\mciteBstWouldAddEndPuncttrue
\mciteSetBstMidEndSepPunct{\mcitedefaultmidpunct}
{\mcitedefaultendpunct}{\mcitedefaultseppunct}\relax
\EndOfBibitem
\bibitem[Gr{\"u}nberg \latin{et~al.}(1982)Gr{\"u}nberg, Cottam, Vach, Mayr, and Camley]{grunberg1982}
Gr{\"u}nberg,~P.; Cottam,~M.; Vach,~W.; Mayr,~C.; Camley,~R. Brillouin scattering of light by spin waves in thin ferromagnetic films. \emph{Journal of Applied Physics} \textbf{1982}, \emph{53}, 2078--2083\relax
\mciteBstWouldAddEndPuncttrue
\mciteSetBstMidEndSepPunct{\mcitedefaultmidpunct}
{\mcitedefaultendpunct}{\mcitedefaultseppunct}\relax
\EndOfBibitem
\bibitem[Dieterle \latin{et~al.}(2019)Dieterle, F{\"o}rster, Stoll, Semisalova, Finizio, Gangwar, Weigand, Noske, F{\"a}hnle, Bykova, \latin{et~al.} others]{dieterle2019}
Dieterle,~G.; F{\"o}rster,~J.; Stoll,~H.; Semisalova,~A.; Finizio,~S.; Gangwar,~A.; Weigand,~M.; Noske,~M.; F{\"a}hnle,~M.; Bykova,~I.; others Coherent excitation of heterosymmetric spin waves with ultrashort wavelengths. \emph{Physical Review Letters} \textbf{2019}, \emph{122}, 117202\relax
\mciteBstWouldAddEndPuncttrue
\mciteSetBstMidEndSepPunct{\mcitedefaultmidpunct}
{\mcitedefaultendpunct}{\mcitedefaultseppunct}\relax
\EndOfBibitem
\bibitem[Trevillian and Tyberkevych(2024)Trevillian, and Tyberkevych]{trevillian2024}
Trevillian,~C.; Tyberkevych,~V. Formation of chirality in propagating spin waves. \emph{npj Spintronics} \textbf{2024}, \emph{2}, 23\relax
\mciteBstWouldAddEndPuncttrue
\mciteSetBstMidEndSepPunct{\mcitedefaultmidpunct}
{\mcitedefaultendpunct}{\mcitedefaultseppunct}\relax
\EndOfBibitem
\bibitem[Heins \latin{et~al.}(2025)Heins, Iurchuk, Gladii, K\"orber, K\'akay, Fassbender, Schultheiss, and Schultheiss]{heins2025}
Heins,~C.; Iurchuk,~V.; Gladii,~O.; K\"orber,~L.; K\'akay,~A.; Fassbender,~J.; Schultheiss,~K.; Schultheiss,~H. Nonreciprocal spin-wave dispersion in magnetic bilayers. \emph{Phys. Rev. B} \textbf{2025}, \emph{111}, 134434\relax
\mciteBstWouldAddEndPuncttrue
\mciteSetBstMidEndSepPunct{\mcitedefaultmidpunct}
{\mcitedefaultendpunct}{\mcitedefaultseppunct}\relax
\EndOfBibitem
\bibitem[Veerakumar and Camley(2006)Veerakumar, and Camley]{veerakumar2006}
Veerakumar,~V.; Camley,~R. Magnon focusing in thin ferromagnetic films. \emph{Physical Review B—Condensed Matter and Materials Physics} \textbf{2006}, \emph{74}, 214401\relax
\mciteBstWouldAddEndPuncttrue
\mciteSetBstMidEndSepPunct{\mcitedefaultmidpunct}
{\mcitedefaultendpunct}{\mcitedefaultseppunct}\relax
\EndOfBibitem
\bibitem[Kim \latin{et~al.}(2016)Kim, Stamps, and Camley]{kim2016}
Kim,~J.-V.; Stamps,~R.~L.; Camley,~R.~E. Spin wave power flow and caustics in ultrathin ferromagnets with the {D}zyaloshinskii-{M}oriya interaction. \emph{Physical Review Letters} \textbf{2016}, \emph{117}, 197204\relax
\mciteBstWouldAddEndPuncttrue
\mciteSetBstMidEndSepPunct{\mcitedefaultmidpunct}
{\mcitedefaultendpunct}{\mcitedefaultseppunct}\relax
\EndOfBibitem
\bibitem[Heussner \latin{et~al.}(2018)Heussner, Nabinger, Fischer, Br{\"a}cher, Serga, Hillebrands, and Pirro]{heussner2018}
Heussner,~F.; Nabinger,~M.; Fischer,~T.; Br{\"a}cher,~T.; Serga,~A.~A.; Hillebrands,~B.; Pirro,~P. Frequency-Division Multiplexing in Magnonic Logic Networks Based on Caustic-Like Spin-Wave Beams. \emph{physica status solidi (RRL)--Rapid Research Letters} \textbf{2018}, \emph{12}, 1800409\relax
\mciteBstWouldAddEndPuncttrue
\mciteSetBstMidEndSepPunct{\mcitedefaultmidpunct}
{\mcitedefaultendpunct}{\mcitedefaultseppunct}\relax
\EndOfBibitem
\bibitem[Prestwood \latin{et~al.}(2025)Prestwood, Barker, Stenning, Freeman, Wei, Kikkawa, Dion, Stoeffler, Henry, Bailleul, Naushad, Griggs, Thomson, Cubukcu, Gartside, Saitoh, Branford, and Kurebayashi]{prestwood2025}
Prestwood,~D. \latin{et~al.}  Spin wave resonance in yttrium iron garnet stripe domains. 2025; \url{https://arxiv.org/abs/2505.08431}\relax
\mciteBstWouldAddEndPuncttrue
\mciteSetBstMidEndSepPunct{\mcitedefaultmidpunct}
{\mcitedefaultendpunct}{\mcitedefaultseppunct}\relax
\EndOfBibitem
\bibitem[Das \latin{et~al.}(2024)Das, Mansell, Flaj\ifmmode~\check{s}\else \v{s}\fi{}man, Yao, van~der Jagt, Chen, Ravelosona, Herrera~Diez, and van Dijken]{das2024}
Das,~S.; Mansell,~R.; Flaj\ifmmode~\check{s}\else \v{s}\fi{}man,~L. c.~v.; Yao,~L.; van~der Jagt,~J.~W.; Chen,~S.; Ravelosona,~D.; Herrera~Diez,~L.; van Dijken,~S. Tuning of perpendicular magnetic anisotropy in Bi-substituted yttrium iron garnet films by ${\mathrm{He}}^{+}$ ion irradiation. \emph{Phys. Rev. Mater.} \textbf{2024}, \emph{8}, 114419\relax
\mciteBstWouldAddEndPuncttrue
\mciteSetBstMidEndSepPunct{\mcitedefaultmidpunct}
{\mcitedefaultendpunct}{\mcitedefaultseppunct}\relax
\EndOfBibitem
\bibitem[Ghising \latin{et~al.}(2017)Ghising, Hossain, and Budhani]{ghising2017}
Ghising,~P.; Hossain,~Z.; Budhani,~R.~C. Stripe magnetic domains in CeY2Fe5O12 (Ce:YIG) epitaxial films. \emph{Applied Physics Letters} \textbf{2017}, \emph{110}, 012406\relax
\mciteBstWouldAddEndPuncttrue
\mciteSetBstMidEndSepPunct{\mcitedefaultmidpunct}
{\mcitedefaultendpunct}{\mcitedefaultseppunct}\relax
\EndOfBibitem
\bibitem[Voltan \latin{et~al.}(2016)Voltan, Cirillo, Snijders, Lahabi, Garc\'{\i}a-Santiago, Hern\'andez, Attanasio, and Aarts]{voltan2016}
Voltan,~S.; Cirillo,~C.; Snijders,~H.~J.; Lahabi,~K.; Garc\'{\i}a-Santiago,~A.; Hern\'andez,~J.~M.; Attanasio,~C.; Aarts,~J. Emergence of the stripe-domain phase in patterned permalloy films. \emph{Phys. Rev. B} \textbf{2016}, \emph{94}, 094406\relax
\mciteBstWouldAddEndPuncttrue
\mciteSetBstMidEndSepPunct{\mcitedefaultmidpunct}
{\mcitedefaultendpunct}{\mcitedefaultseppunct}\relax
\EndOfBibitem
\bibitem[Sebastian \latin{et~al.}(2015)Sebastian, Schultheiss, Obry, Hillebrands, and Schultheiss]{sebastian2015}
Sebastian,~T.; Schultheiss,~K.; Obry,~B.; Hillebrands,~B.; Schultheiss,~H. Micro-focused Brillouin light scattering: imaging spin waves at the nanoscale. \emph{Frontiers in Physics} \textbf{2015}, \emph{3}, 35\relax
\mciteBstWouldAddEndPuncttrue
\mciteSetBstMidEndSepPunct{\mcitedefaultmidpunct}
{\mcitedefaultendpunct}{\mcitedefaultseppunct}\relax
\EndOfBibitem
\bibitem[Perzlmaier \latin{et~al.}(2008)Perzlmaier, Woltersdorf, and Back]{perzlmaier2008}
Perzlmaier,~K.; Woltersdorf,~G.; Back,~C.~H. Observation of the propagation and interference of spin waves in ferromagnetic thin films. \emph{Physical Review B—Condensed Matter and Materials Physics} \textbf{2008}, \emph{77}, 054425\relax
\mciteBstWouldAddEndPuncttrue
\mciteSetBstMidEndSepPunct{\mcitedefaultmidpunct}
{\mcitedefaultendpunct}{\mcitedefaultseppunct}\relax
\EndOfBibitem
\bibitem[Wojewoda \latin{et~al.}(2023)Wojewoda, Ligmajer, Hrto{\v{n}}, Kl{\'\i}ma, Dhankhar, Dav{\'\i}dkov{\'a}, Sta{\v{n}}o, Holobr{\'a}d{\'e}k, Kr{\v{c}}ma, Zl{\'a}mal, {\v{S}}ikola, and Urb{\'a}nek]{wojewoda2023}
Wojewoda,~O.; Ligmajer,~F.; Hrto{\v{n}},~M.; Kl{\'\i}ma,~J.; Dhankhar,~M.; Dav{\'\i}dkov{\'a},~K.; Sta{\v{n}}o,~M.; Holobr{\'a}d{\'e}k,~J.; Kr{\v{c}}ma,~J.; Zl{\'a}mal,~J.; {\v{S}}ikola,~T.; Urb{\'a}nek,~M. Observing high-k magnons with Mie-resonance-enhanced Brillouin light scattering. \emph{Communications Physics} \textbf{2023}, \emph{6}, 94\relax
\mciteBstWouldAddEndPuncttrue
\mciteSetBstMidEndSepPunct{\mcitedefaultmidpunct}
{\mcitedefaultendpunct}{\mcitedefaultseppunct}\relax
\EndOfBibitem
\bibitem[Gr{\"a}fe \latin{et~al.}(2019)Gr{\"a}fe, Weigand, Van~Waeyenberge, Gangwar, Gro{\ss}, Lisiecki, Rychly, Stoll, Tr{\"a}ger, F{\"o}rster, Stobiecki, Dubowik, K{\l}os, Krawczyk, Back, Goering, and Sch{\"u}tz]{grafe2019}
Gr{\"a}fe,~J. \latin{et~al.}  Visualizing nanoscale spin waves using MAXYMUS. Spintronics XII. Bellingham, Washington, 2019; p 1109025\relax
\mciteBstWouldAddEndPuncttrue
\mciteSetBstMidEndSepPunct{\mcitedefaultmidpunct}
{\mcitedefaultendpunct}{\mcitedefaultseppunct}\relax
\EndOfBibitem
\bibitem[Wittrock \latin{et~al.}(2025)Wittrock, Klose, Perna, Baumgaertl, Mucchietto, Schneider, Fuchs, Deinhart, Karaman, Grundler, Eisebitt, Pfau, and Schick]{wittrock2025}
Wittrock,~S.; Klose,~C.; Perna,~S.; Baumgaertl,~K.; Mucchietto,~A.; Schneider,~M.; Fuchs,~J.; Deinhart,~V.; Karaman,~T.; Grundler,~D.; Eisebitt,~S.; Pfau,~B.; Schick,~D. Soft-X-ray momentum microscopy of nonlinear magnon interactions below 100-nm wavelength. 2025\relax
\mciteBstWouldAddEndPuncttrue
\mciteSetBstMidEndSepPunct{\mcitedefaultmidpunct}
{\mcitedefaultendpunct}{\mcitedefaultseppunct}\relax
\EndOfBibitem
\bibitem[Va\ifmmode~\check{n}\else \v{n}\fi{}atka \latin{et~al.}(2021)Va\ifmmode~\check{n}\else \v{n}\fi{}atka, Szulc, Wojewoda, Dubs, Chumak, Krawczyk, Dobrovolskiy, K\l{}os, and Urb\'anek]{Szulc2021}
Va\ifmmode~\check{n}\else \v{n}\fi{}atka,~M.; Szulc,~K.; Wojewoda,~O.; Dubs,~C.; Chumak,~A.~V.; Krawczyk,~M.; Dobrovolskiy,~O.~V.; K\l{}os,~J.~W.; Urb\'anek,~M. Spin-Wave Dispersion Measurement by Variable-Gap Propagating Spin-Wave Spectroscopy. \emph{Phys. Rev. Appl.} \textbf{2021}, \emph{16}, 054033\relax
\mciteBstWouldAddEndPuncttrue
\mciteSetBstMidEndSepPunct{\mcitedefaultmidpunct}
{\mcitedefaultendpunct}{\mcitedefaultseppunct}\relax
\EndOfBibitem
\bibitem[Gruszecki \latin{et~al.}(2022)Gruszecki, Guslienko, Lyubchanskii, and Krawczyk]{Gruszecki2022}
Gruszecki,~P.; Guslienko,~K.~Y.; Lyubchanskii,~I.~L.; Krawczyk,~M. Inelastic Spin-Wave Beam Scattering by Edge-Localized Spin Waves in a Ferromagnetic Thin Film. \emph{Phys. Rev. Appl.} \textbf{2022}, \emph{17}, 044038\relax
\mciteBstWouldAddEndPuncttrue
\mciteSetBstMidEndSepPunct{\mcitedefaultmidpunct}
{\mcitedefaultendpunct}{\mcitedefaultseppunct}\relax
\EndOfBibitem
\bibitem[Sobucki \latin{et~al.}(2023)Sobucki, {\'S}migaj, Graczyk, Krawczyk, and Gruszecki]{Sobucki2023}
Sobucki,~K.; {\'S}migaj,~W.; Graczyk,~P.; Krawczyk,~M.; Gruszecki,~P. Magnon-Optic Effects with Spin-Wave Leaky Modes: Tunable Goos-{H}\"anchen Shift and {W}ood’s Anomaly. \emph{Nano Letters} \textbf{2023}, \emph{23}, 6979--6984\relax
\mciteBstWouldAddEndPuncttrue
\mciteSetBstMidEndSepPunct{\mcitedefaultmidpunct}
{\mcitedefaultendpunct}{\mcitedefaultseppunct}\relax
\EndOfBibitem
\end{mcitethebibliography}

\newpage

\appendix
\renewcommand{\thesection}{\Alph{section}} 
\renewcommand{\thesubsection}{\thesection.\arabic{subsection}} 
\numberwithin{equation}{section} 

\renewcommand{\theequation}{A\arabic{equation}}
\setcounter{equation}{0}  

\renewcommand{\thefigure}{A\arabic{figure}}
\setcounter{figure}{0}  




\FloatBarrier

\end{document}